\documentclass[%
reprint,
superscriptaddress,
twocolumn,
 amsmath,amssymb,
 aps,
 prx,
]{revtex4-2}

\usepackage{graphicx}% Include figure files
\usepackage{dcolumn}% Align table columns on decimal point
\usepackage{bm}% bold math
\usepackage{comment}
\usepackage{hyperref}% add hypertext capabilities
\usepackage{xcolor}
\renewcommand{\selectlanguage}[1]{}

\begin{document}

\title{Cell bulging and extrusion
in a three-dimensional bubbly vertex model\\
for curved epithelial sheets}

\author{Oliver M. Drozdowski}
\affiliation{Institute for Theoretical Physics, Heidelberg University, 69120 Heidelberg, Germany}
\affiliation{BioQuant, Heidelberg University, 69120 Heidelberg, Germany}
\affiliation{Max Planck School Matter to Life, Heidelberg University, 69120 Heidelberg, Germany}
\affiliation{Institute of Science and Technology Austria, 3400 Klosterneuburg, Austria}
\author{B\"u\c{s}ra Kocame\c{s}e}
\affiliation{BioQuant, Heidelberg University, 69120 Heidelberg, Germany}
\affiliation{Division Signaling and Functional Genomics, German Cancer Research Center (DKFZ), 69120 Heidelberg, Germany}
\author{Kim E. Boonekamp}
\affiliation{BioQuant, Heidelberg University, 69120 Heidelberg, Germany}
\affiliation{Division Signaling and Functional Genomics, German Cancer Research Center (DKFZ), 69120 Heidelberg, Germany}
\author{Michael Boutros}
\affiliation{BioQuant, Heidelberg University, 69120 Heidelberg, Germany}
\affiliation{Max Planck School Matter to Life, Heidelberg University, 69120 Heidelberg, Germany}
\affiliation{Division Signaling and Functional Genomics, German Cancer Research Center (DKFZ), 69120 Heidelberg, Germany}
\affiliation{Department of Molecular Human Genetics, Medical Faculty Heidelberg, Heidelberg University, 69120 Heidelberg, Germany}
\author{Ulrich S. Schwarz}
\email{Corresponding author: schwarz@thphys.uni-heidelberg.de}
\affiliation{Institute for Theoretical Physics, Heidelberg University, 69120 Heidelberg, Germany}
\affiliation{BioQuant, Heidelberg University, 69120 Heidelberg, Germany}
\affiliation{Max Planck School Matter to Life, Heidelberg University, 69120 Heidelberg, Germany}

\date{\today}

\begin{abstract}
Cell extrusion is an essential mechanism for controlling cell density in epithelial tissues. Another
essential element of epithelia is curvature, which is required to achieve complex shapes, like in the
lung or intestine. Here we introduce a three-dimensional bubbly vertex model to study the interplay
between extrusion and curvature. We find a generic cellular bulging instability at topological defects
which is much stronger than for standard vertex models.
Analyzing cell shapes in three-dimensional imaging data of spherical mouse colon organoids, we 
infer that pentagonal cells have an increased basal interfacial tension, 
suggesting that cells at topological defects react to the different
force conditions. Using the 
bubbly vertex model,  we show that 
such basal tensions stabilize against the predicted instability and result in better cell shape control 
than tissue-scale mechanisms such as lumen pressure and spontaneous curvature. Our theory suggests that
epithelial curvature naturally leads to bulged and extrusion-like cell shapes because the interfacial curvature of individual cells
at the defects strongly amplifies
buckling effected by tissue-scale topological defects in elastic sheets.  Our results highlight the complex interplay 
of forces across scales in three-dimensional tissue organization.
\end{abstract}

\maketitle

\section{Introduction}

Epithelial tissues are close-packed and sealed cell 
layers that form the
barriers between different compartments in the body,
for example in the lung or the intestine. 
In order to regulate their cell density, they
have to achieve a balance between cell division
and cell death. Recently it has become clear that
both processes are strongly connected to mechanics:
cells divide more if they are
stretched by neighboring cells at low cell density, and 
they stop dividing
and die if the epithelium becomes too crowded \cite{Eisenhoffer_Nat12_Live_Cell_extrusion_crowding, Gudipaty_Nat17_Mechanica_stretch_triggers_division_piezo1,GudipatyRosenblatt_SemCellDevB17_Cell_extrusion_pathways}. 
In order to make sure that cells 
leave the tissue without disrupting its barrier
function, they are usually expelled by extrusion, 
which means they are squeezed out of the epithelium
without the generation of any voids \cite{Kocgozlu2016_CurrBiol_Cell_packing_extrusion_modes,Le_NatComms21_Apoptotic_cell_extrusion_heterogeneous_actin_organization}. 
Extrusion is especially prominent in the intestine, 
where the epithelium renews itself every 3-5 days \cite{Clevers_AnnRevPhysiol09_Self_renewal_intestinal_epithelium}.
Here cells proliferate in the crypt regions and then
move in a constant stream to the tip of villi, 
where they are extruded \cite{PerezGonzales_Trepat_CurrOpGenDev22_Mechanobiology_intestinal_epithelium}. 
Very recently, it has been shown that most cells extruded in the 
villus regions are still alive, which means that they mechanically
do not differ much from their neighbors \cite{Krueger_Clevers_Sciene_2025_Extrusion}.
The exact mechanisms of extrusion are not clear, but one important element
seems to be the presence of specialized actin structures on the lateral
and basal sides \cite{GudipatyRosenblatt_SemCellDevB17_Cell_extrusion_pathways,Krueger_Clevers_Sciene_2025_Extrusion}.
For flat model epithelia, it has been shown that cell extrusion
occurs preferentially at topological defects in the nematic orientation field
of the cells \cite{Saw_Nat_2017_topological_defects_extrusion, FadulRosenblatt_CurrOpCellBiol18_Forces_fates_extruding_cells, Chen_JCellSci18_Review_mechanical_forces_monolayers}.

In general, cells in non-stratified epithelia tend to have 
hexagonal order, because this is the close-packed optimum in
two dimensions, similar to the situation in sphere
packings and foams. Hexagonal order is also dominant in curved epithelial sheets
\cite{Tang_NatPhys_2022_Hexagonal_islands_curved_epithelia, Hoffmann_Giomi_SciAdv_22_Defect_mediated_morphogenesis, Eckert_NatComms23_Hexanematic_epithelial_crossover_adhesion_density, Armengol-Collado_NatPhys2023_Hexanematic_crossover_epithelia}. 
However, now some pentagonal defects are required to generate
curvature \cite{Hoffmann_Giomi_SciAdv_22_Defect_mediated_morphogenesis}. 
Due to Euler's polyhedron theorem, 12 pentagonal defects
are required to achieve the topology of a sphere,
with the possibility of additional pentagonal-heptagonal pairs.
Thus topological defects occur naturally in 
curved epithelia, while in flat cell layers,
they arise rather in a stochastic manner from the movement
of cells. Because epithelia are often curved due to their biological function, for example in the
crypts and villi of the intestine, the question arises if they might use the naturally occurring topological defects to promote cell extrusion. 
While previous theoretical models have addressed such extrusion in a flat configuration focusing on fluid-like epithelia \cite{Monfared_elife23_Mechanics_topological_routes_cell_elimination} or neglecting the three-dimensional nature of the problem \cite{Bielmeier_CurrBiol16_Interface_contractility_cell_elimination}, the coupling of curvature, defects, and shape instabilities has hardly been addressed before. 
Defects in such a setting have been linked to extrusion in flat epithelia on a substrate in phase field models of tissues \cite{Monfared_elife23_Mechanics_topological_routes_cell_elimination} and to shape instabilities in vertex models \cite{Okuda_BPJ20_Mechanical_instability_epithelial_monolayers_extrusion}, but the underlying mechanisms and their relation
to curvature have not been addressed yet.

Recently, we have reported on a morphological instability at topological disclination defects in curved hexagonal epithelia in the three-dimensional vertex model (VM) \cite{Drozdowski_PRR24_Topological_defects_VM_shells}.
This buckling instability arises because
spherical epithelia are effectively elastic shells,
similar to viral capsids \cite{Lidmar_PRE_2003_icosahedral_instability_shells, Nguyen_PRE05_viral_capsid_spontaneous_curvature} or buckyballs \cite{Witten_EuPhysLett93_Fullerene_Ball_icosahedral_instability},
for which such an instability is known since the pioneering
work by Seung and Nelson \cite{Seung_88_PRL_Conical_instability_disclination}.
However, the standard VM does not consider that cells
might adapt their own interfacial curvature to
their mechanical environment.
Here we investigate this aspect by using the theoretical bubbly vertex model (BVM) 
and by analyzing experimentally observed cell shapes in spherical mouse colon organoids. 
In contrast to the normal VM, which describes multicellular systems with
flat interfaces, in the BVM the interfaces are curved \cite{Ishimoto_PRE14_Bubbly_vertex_dynamics, Boromand_PRL18_Jamming_deformable_polygons, runser_NatCompSci24_SimuCell3D_bubbly_vertex_with_nucleus_and_bending, Staddon_2024_Curved_edges_vm_fluidity, Arroyo2025_active_gel_bubbly_vertex_model}.
Such interfacial curvature leverages the three-dimensional nature of the cellular system and 
allows for different mechanisms than the usual VM-approach to epithelia. On the
experimental side, colon organoids are particularly suited to address the coupling of mechanics, cell shape and topology, as they form spherical epithelial monolayers. They are derived from stem cells gathered from colonic crypts, recapitulate developmental processes of the colon and thus serve as important model system for development and disease \cite{Sprangers2021_Review_organoids}.

We start in Section~\ref{sec:bvm} by showing with computer simulations
that the curved interfaces in the BVM strongly
amplify the buckling at defects in curved sheets if
compared to the standard VM with flat interfaces.
Through segmentation of 3D microscopy data of mouse colon organoids with the subsequent inference of interfacial tensions, we find that
cellular shapes are more regular in spherical organoids than theoretically expected.
We also infer that outer (basal) tension are increased at pentagonal cells, suggesting that cells react to the changed force conditions
at topological defects.
In the BVM, we find such increased basal tensions to stabilize against the predicted instability and to yield full control over cellular bulging. 
To understand the underlying instability theoretically, 
we consider the representative case of a shell with icosahedral symmetry in which interfacial curvature 
leads to a significant lowering of a previously described icosahedral buckling threshold \cite{Drozdowski_PRR24_Topological_defects_VM_shells, Lidmar_PRE_2003_icosahedral_instability_shells, Seung_88_PRL_Conical_instability_disclination}.
In Section~\ref{sec:continuum} we develop a continuum description of a defect cell in a mean field tissue,
allowing us to extend our model to describe and study extrusion-like cell geometries.
We find that the possibility of curved interfaces results in a local decrease of the saddle splay modulus, thus leading to localization of Gaussian curvature in the defect.
We find that lowering the apico-basal surface tension and a supracellular actin ring may initiate bulging, while only the former allows for a transition of a partially extruded geometry toward full extrusion.
Our model predicts a lower energy barrier for extrusion at topological (pentagonal) defects.
We close in Section~\ref{sec:discussion} with a discussion of our results.

\section{Bubbly Vertex Model and bulging instability at topological defects}\label{sec:bvm}

\subsection{Definition of bubbly vertex model (BVM)}

\begin{figure}[!t]
	\centering
	\includegraphics[width=\columnwidth]{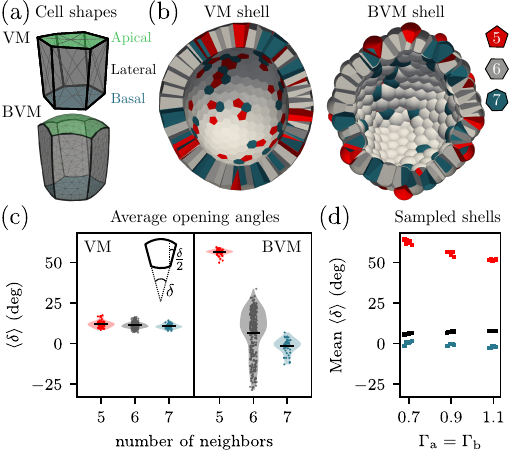}
	\caption{Cell bulging at topological defects in the bubbly vertex model.
 (a) Schematic depiction of the vertex model (VM) representation of a cell with apical, basal and lateral faces via an interpolation point of the polygonal faces, and of the bubbly vertex model (BVM) representation with curved interfaces.
 (b) Spherical VM and BVM shell after energy minimization with random initialization of cell centers. BVM shell is after minimization in Surface Evolver. Number of cell neighbors is color-coded.
 (c) Opening angles of the cells by neighbor number for the VM and BVM, where the half angle is determined for each lateral interface (inset) and then averaged over lateral interfaces. Violin plots show distribution with the horizontal line marking the mean value. Tensions $\Gamma_\mathrm{a}=\Gamma_\mathrm{b}=0.9$, $\Gamma_\mathrm{l}=1$.
(d) Mean opening angles of cells with different neighbor numbers (color-coded; red pentagons, black hexagons, blue heptagons) 
 in different BVM-shells with random initial cell configurations ($N=6$) and varying apical and basal tensions $\Gamma_\mathrm{a}=\Gamma_\mathrm{b}$ and $\Gamma_\mathrm{l}=1$.
 All shells with $400$ cells. 
 }
\label{fig:fig_shells_pentagonal_collapse}
\end{figure}

For the description of tissue mechanics and shape, we adopt the usual tension-based approach. 
We start from the regular three-dimensional vertex model for epithelial monolayers 
in which individual cells are described as bounded volumes with one apical, one basal and multiple lateral faces as boundaries. 
While in the classical vertex model these boundary faces are assumed to be flat by triangulating them with a passive interpolation point \cite{Krajnc_PRE18_VM_fluidization,Drozdowski_PRR24_Topological_defects_VM_shells}, 
we explicitly want to consider the effect of possible interface curvature on the mechanics
and thus introduce the bubbly vertex model, in which the interfaces
are allowed to curve \cite{Ishimoto_PRE14_Bubbly_vertex_dynamics, Boromand_PRL18_Jamming_deformable_polygons, runser_NatCompSci24_SimuCell3D_bubbly_vertex_with_nucleus_and_bending, Staddon_2024_Curved_edges_vm_fluidity, Arroyo2025_active_gel_bubbly_vertex_model}, cf.~Fig.~\ref{fig:fig_shells_pentagonal_collapse}(a).

Each cell $i$ has one apical area $A_{a,i}$, one basal area $A_{b,i}$ and the integrated lateral area $A_{l,i}$, corresponding to the areas of the boundary faces. 
We assume the faces to have surface tensions 
$\Gamma_\mathrm{a}$, $\Gamma_\mathrm{b}$ and $\Gamma_\mathrm{l}$ for the apical, basal, and lateral faces, respectively, 
and that their values are the same for all cells (no dependance on $i$).
The total energy of the tissue is then described by the sum of the individual surface area contributions
\begin{equation}\label{eq:continuum_bubbly_vertex_energy}
    E = \sum_{i} \left( \Gamma_{a} A_{a,i} + \Gamma_{b} A_{b,i} + \frac{1}{2} \Gamma_{l} A_{l,i} \right),
\end{equation}
where the factor $1/2$ accounts for double counting. 
The apical and basal network topologies are identical, i.e., 
both apical and basal sides have the same number of adjacent apical and basal faces, respectively, and lateral faces are quadrilateral. 
We assume each cell to have a fixed volume $V_\mathrm{cell}$. 
We non-dimensionalize the surface tension via $\Gamma_l$ and the volume via $V_\mathrm{cell}$, i.e., we set $\Gamma_l=1$ and $V_\mathrm{cell}=1$.

Due to the fluid viscoelastic behavior of the cell cortex \cite{Janshoff_BioChem21_Review_viscoelasticity_epithelial_cells, Cordes_PRL20_Prestress_area_compressibility_actin_cortex_viscoelastic_response_cells}, we neglect elastic in-plane contributions of the interfaces, as these will relax over sufficiently long time scales.
For now, we also do not consider a bending energy contribution \cite{runser_NatCompSci24_SimuCell3D_bubbly_vertex_with_nucleus_and_bending}, making use of the analogy to soap films, which are purely driven by surface tensions.
This can be justified by the small thickness of the cortex, which will lead to small bending rigidities in the case of large interfacial tensions. 
We will relax this assumption later.

To differentiate between the model with and without membrane curvature, we call the former vertex model (VM) and the latter bubbly vertex model (BVM) throughout the rest of this article. 

\subsection{Instability at topological defects in the BVM}

We consider a spherical epithelium with a randomized starting configuration. 
For simulations we use \textit{OrganoidChaste} \cite{OrganoidChaste}, a custom-written package for Chaste \cite{Chaste_Software_2020} implementing a three-dimensional vertex model for monolayers, which was used in Ref.~\cite{Drozdowski_PRR24_Topological_defects_VM_shells} and is described in Appendix~\ref{app:organoidchaste}.
Cell centers are created using the random sequential adsorption mechanism \cite{Roshal_PRE23_Topological_properties_monolayer_random_sequential_adsorption}
and then a Voronoi tesselation on the sphere to determine cell shape.
Using a rate of active topological T1-transitions which is slowly reduced to zero, the energy is minimized to obtain a spherical starting configuration.
This process is performed in the VM to obtain shapes with low energy
and a representative minimal energy configuration 
is shown in Fig.~\ref{fig:fig_shells_pentagonal_collapse}(b). 
This topology is then fixed for the comparison between VM and BVM shapes, as here we do not consider 
tissue dynamics on a larger time scale.

Next we minimized the energy in the BVM by using the VM configuration as starting configuration in
the standard software \textit{Surface Evolver} for energy minimization of surfaces under tension \cite{Brakke_92_Surface_Evolver}.
Again we minimize at constant cell volume and get the  
result shown in Fig.~\ref{fig:fig_shells_pentagonal_collapse}(b). As shown by the color code, for both 
the VM and the BVM, pentagons, hexagons and
heptagons are distributed all over the surface.
The spherical configuration necessitates curvature screening \cite{GarciaAguilar_Giomi_PRE20_dislocation_screening} through additional pairs of topological defects (pentagon-heptagon pairs) 
compared to an icosahedral configuration of a hexagonal lattice with 12 pentagons.
While topological disclination defects indeed seem to be screened by the 
distribution of additional defect pairs in the VM, 
leading to an approximately spherical configuration, 
we observe an interesting instability in the BVM.
Pentagonal cells tend to bulge outward, minimizing their luminal interfacial area, 
despite the fact that we do not observe a tissue-scale buckling instability.
Fig.~\ref{fig:fig_shells_pentagonal_collapse}(c) demonstrates the dramatic
difference between the two models: the opening angles of pentagonal cells are 
much larger in the BVM than in the VM, agreeing with the visual 
impression when comparing shells in Fig.~\ref{fig:fig_shells_pentagonal_collapse}(b).
In addition, we also see that the distribution of opening angles in the BVM is much 
broader with some cells even bending in the opposite direction. We conclude that the 
BVM allows for more adaptation and variability in cell shape than the classical VM,
with a particularly strong effect on the pentagonal defects. We verified the generality of this result, by sampling many random configurations for shells with the same number of cells and for different apico-basal tensions $\Gamma_\mathrm{a}=\Gamma_\mathrm{b}$, effectively considering different shell sizes with respect to the monolayer thickness, cf.~Fig.~\ref{fig:fig_shells_pentagonal_collapse}(d).

\subsection{Cell shapes at topological defects in mouse colon organoids}
\begin{figure*}[!t]
	\centering
	\includegraphics[width=6in]{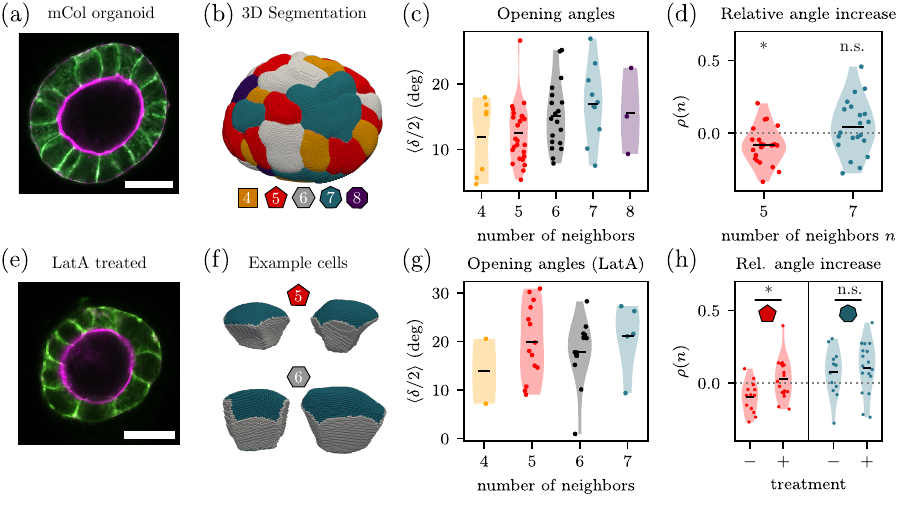}
	\caption{ Experimentally observed cell shapes at topological defects in mouse colon (mCol) organoids. 
 (a) Microscopy image of spherical mouse colon organoid. Green is membrane ($\beta$-catenin) and magenta is F-actin. The luminal side is apical (high actin) and the outer side basal (facing the matrix).
  (b) The organoid is segmented in 3D and the number of neighbors (color-coded) is determined from the basal polygonal edges.
  (c) For this organoid the opening angles $\delta/2$, averaged over all lateral interfaces, vary for different neighbor numbers. Each dot represents one cell, the bars depict the means.
  (d) The relative angle increase at pentagonal and heptagonal cells, compared to hexagonal cells. Every data point is one organoid ($N=22$).
  (e) Microscopy image of organoid treated with Latrunculin A (LatA).
  (f) Examples of pentagonal and hexagonal cells as segmented in 3D.
  (g) Opening angles for organoid treated with LatA.
  (h) Comparing the relative angle increase $\rho$ from wildtype ($N=13$) and LatA-treated samples ($N=17$),
  we find larger effects at topological defects with LatA.
 Scale bars in (a) and (e) are $20\,\mu\mathrm{m}$. Stars indicate: $*\ p<0.05$, $**\ p<0.01$, $***\ p<0.001$. 
 Tests against $0$ in (d) with Wilcoxon and against each other in (h) with Mann-Whitney-U tests.
 }
\label{fig:exp_mcol}
\end{figure*}

Mouse colon (mCol) organoids are spherical epithelia derived from single stem cells \cite{Sprangers2021_Review_organoids, Sato2011_Gastroenterology_mcol_organoids}.
Fig.~\ref{fig:exp_mcol}(a) shows a typical microscopy image.
We have grown these organoids in extracellular matrix (basement membrane extract) from mechanically dissociated (split) 
closed organoids and fixed, stained and 
imaged them at day 5 post-splitting. 
The luminal (inner) side of the organoids corresponds to the apical side of the cells, as demonstrated by
the actin staining in Fig.~\ref{fig:exp_mcol}(a).
The outer side corresponds to the basal side of the cells and has been growing against the matrix.
The three-dimensional images were segmented, as shown in Fig.~\ref{fig:exp_mcol}(b), 
using a previously described pipeline that
we have developed in the context of force inference \cite{Drozdowski_Boonekamp_2025_Force_inference} 
(see Appendices~\ref{app:organoid_exp_analysis} and \ref{app:segmentaion_force_inference} for details). 
Similarly to the BVM, we determined the (half) cellular opening angles $\delta/2$ of the mCol cells 
as a function of their number of neighbors. Contrary to what we expect from the BVM, 
we find that in some organoids the opening angles of pentagonal cells are smaller than for hexagons
and vice versa for heptagons (Fig.~\ref{fig:exp_mcol}(c) for one organoid).
To allow for a comparison across organoids, we determined the relative increase of cell-averaged pentagonal ($n=5$) and heptagonal ($n=7$) opening angles, 
$\langle\delta/2\rangle_n$, compared to the angles averaged over hexagonal cells, $\langle\delta/2\rangle_6$, i.e.
\begin{equation}
\rho(n) = \frac{\langle\delta/2\rangle_n - \langle \delta/2\rangle_6}{\langle \delta/2\rangle_6}.
\end{equation}
The result is shown in Fig.~\ref{fig:exp_mcol}(d).
While we find this angle increase in pentagons to be larger than zero in some organoids, 
consistent with the bulging instability, 
it is significantly lower than zero on the level of many organoids.
In heptagonal cells, this difference is not significant.

The experimental data show that cell shape is indeed significantly different at topological defects, 
albeit oppositely to the bulging instability predicted by the BVM.
As the actomyosin systems has a stabilizing function through both the formation of an apical (luminal) actin belt and interfacial contractility, we treated the mCol organoids with Latrunculin A, 
an inhibitor of actin polymerization, effectively leading to actin depolymerization.
While overall organoid morphology was not disrupted (Fig.~\ref{fig:exp_mcol}(e)),
we found a signature of the bulging instability in multiple treated organoids (Fig.~\ref{fig:exp_mcol}(f,g)).
Indeed, looking at the relative angle increase $\rho(n)$, we found increased angles at pentagonal cells in treated organoid populations, 
hinting at a potentially stabilizing role of the actomyosin system against the bulging instability at disclination defects
(Fig.~\ref{fig:exp_mcol}(h)).
In heptagonal cells we do not see a significant difference between treated and untreated organoids, which could be explained by the smaller relevance of this instability here due to the lack of saddle-like structures in these spherical organoids, cf.~Fig.~\ref{fig:fig_shells_pentagonal_collapse}(c).

\begin{figure}[!t]
	\centering
	\includegraphics[width=\columnwidth]{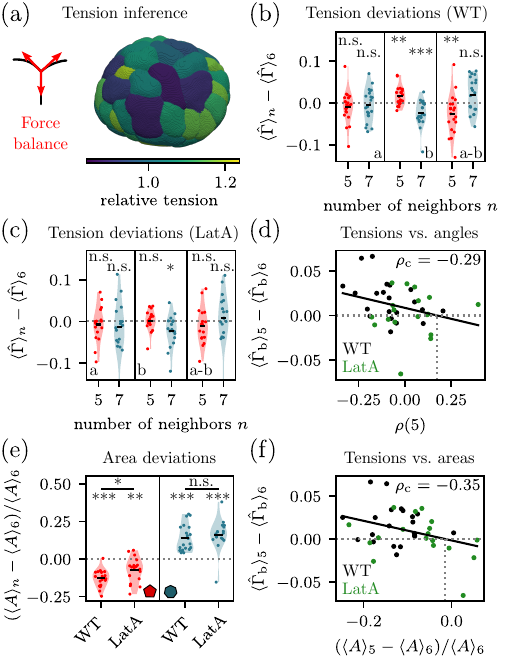}
	\caption{Surface tension inference in mouse colon (mCol) organoids.
 (a) Assuming force balance, the relative tensions can be inferred from the dihedral angles of the interfaces. 
 For the reconstructed organoid the outer (basal) tensions are shown.
(b,c) For each wildtype (WT) (b) and Latrunculin A treated (LatA) (c) organoid we determine the difference of the averaged inferred tensions $\hat{\Gamma}$ for $n$-gons from hexagons for apical (a) and basal (b) interfaces. The averaged cell-specific apico-basal tension difference is also shown (a-b).
(d) Pentagonal basal tension deviations versus relative angle increase of pentagons compared to hexagons $\rho(5)$ for WT and LatA samples. 
Solid line is linear regression, which intersects a tension deviation of $0.0$ at $\rho(5)\approx0.17$. 
(e) Relative mid-plane area deviations of pentagons (red) and heptagons (blue) from hexagons for WT and LatA organoids.
(f) Pentagonal basal tension deviations versus relative area deviations of pentagons for WT and LatA samples. Solid line is linear regression, which intersects a tension deviation of 0.0 at $\left(\langle A\rangle_5-\langle A \rangle_6\right)/\langle{A}\rangle_6=-0.015$.
Pearson correlation coefficient in (d,f) is $\rho_\mathrm{c}$.
Stars indicate: $*\ p<0.05$, $**\ p<0.01$, $***\ p<0.001$. 
 Tests against $0$ in (b,c,e) with Wilcoxon test and against each other in (e) with Mann-Whitney-U-test.}
\label{fig:fig_exp_tens}
\end{figure}

Cellullar shapes display more regular opening angles than predicted by the BVM, suggesting more 
controlled force generation. In particular, mechanosensing and an active compensation mechanism may stabilize 
shape.  In order to test this, we determined the interfacial tensions of the individual apical, basal and lateral interfaces, 
denoted with $\hat{\Gamma}$, from the microscopy images. 
Assuming mechanical equilibrium, interfacial tensions lead to forces normal to tri-interfacial junctions \cite{Drozdowski_Boonekamp_2025_Force_inference, Veldhuis_PTRSB17_Force_inference_image_stacks, Xu2018_PONE_3d_force_inference_pointcloud, Roffay_Dev21_Review_force_inference, Ichbiah_NatMeth23_Embryo_mechanics_cartography}.
Using our recently developed force inference method \cite{Drozdowski_Boonekamp_2025_Force_inference}, we inferred the underlying relative tensions from the interfacial dihedral angles in the segmented images
(cf.~Fig.~\ref{fig:fig_exp_tens}(a), see Appendix \ref{app:organoid_exp_analysis} for details).
Figures~\ref{fig:fig_exp_tens}(b,c) display the different average tensions of different interface types at pentagonal and heptagonal cells, $\langle\hat{\Gamma}\rangle_n$, from the respective averages in hexagonal cells, $\langle\hat{\Gamma}\rangle_6$ 
for wildtype (WT) and Latrunculin A (LatA) treated cells.
Apical tensions are not different in pentagonal and heptagonal cells, but basal tensions are higher (lower) at pentagonal (heptagonal) cells for WT cells.
Treatment of LatA lowers this basal tension, as could be expected from the increased opening angles.
The basal tension increase at pentagonal cells is negatively correlated with the relative angle increase at pentagonal cells (correlation $\rho_\mathrm{c}=-0.29$ with p-value $p=0.07$), supporting the notion of a complex balance between opening angles and basal tension.
Linear regression on both the WT and LatA data suggests a relative angle increase of approximately $17\%$ if the pentagonal basal tension is not different from the hexagonal one (Fig.~\ref{fig:fig_exp_tens}(d)).
This indeed hints at an active compensation effect being responsible for the shape difference. 

To explore potential mechanisms by which cells could adapt dynamically to their topological neighborhood, we determined the mid-plane areas as the ratio of cellular volume and height, $V_\mathrm{cell}/h$ (cf.~Fig.~\ref{fig:fig_exp_tens}(e)).
Similarly to Lewi's law in two-dimensional epithelia and foams which states that area increases with the number of neighbors \cite{Lewis_AnR1928_lewis_law_cucumbers,Szeto_PhysA95_Lewis_law_soap_froth,Gomez-Galvez_Dev21_complex_3d_organization_epithelia_vm_review}, we find approximately $10\%$ smaller pentagonal and $15\%$ larger heptagonal areas.
This enables a direct feedback of tissue topology to the cell and could be linked to mechanosensing, similarly as in density sensing in intestinal epithelia  \cite{Eisenhoffer_Nat12_Live_Cell_extrusion_crowding,Fernandez-Sanchez_Nat2015_beta_catenin_tumor_growth_pressure_colon}.
Looking at the relationship of pentagonal basal tensions and areas, we find a negative correlation $\rho_\mathrm{c}=-0.34$ (p-value $p=0.03$): for smaller average pentagonal areas we find increased basal tensions (Fig.~\ref{fig:fig_exp_tens}(f)). 
Linear regression suggests that this tension increase vanishes as areas become comparable to hexagonal areas.
This is consistent with an area-driven feedback via mechanosensing facilitating a topology-dependent stabilization.
Nonetheless, the underlying bulging instability is not as prominent in mouse colon organoids, implying the need for stabilization mechanisms also in our theoretical description.

\subsection{Predictions of BVM for different shape stabilization mechanisms}
\begin{figure*}[!t]
	\centering
	\includegraphics[width=\textwidth]{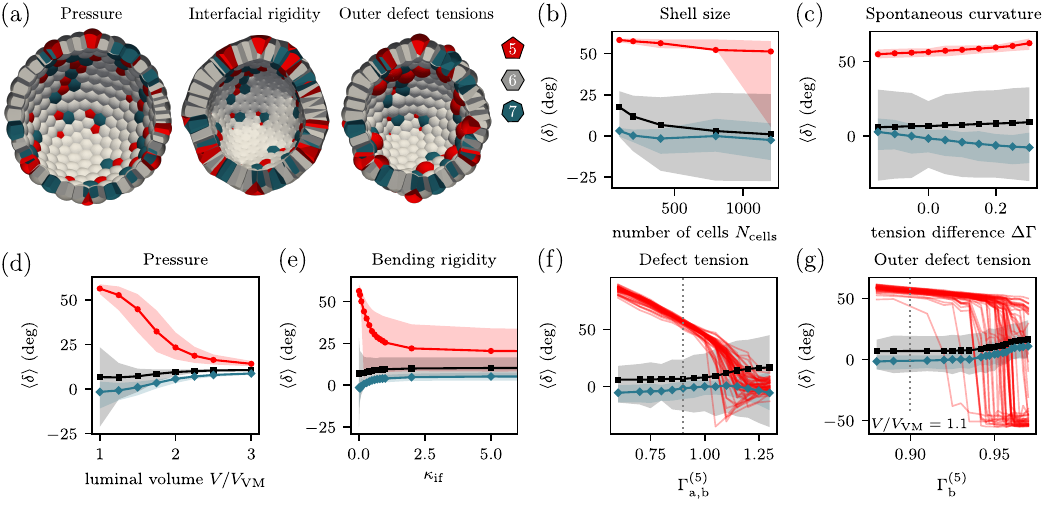}
	\caption{Modulation of instability via different mechanisms explored through the BVM.
(a) The shell from Fig.~\ref{fig:fig_shells_pentagonal_collapse} 
simulated with additional effects: 
pressurized with $V/V_\mathrm{VM}=2.0$, 
rigidified with $\kappa_\mathrm{if}=0.7$,
and pentagonal basal (outer) defect tension of $\Gamma_\mathrm{b}^{(5)}=0.957$. 
 (b) Opening angles of pentagonal (red), hexagonal (black) and heptagonal cells (blue) as a function of the number of simulated cells.
  (c) Opening angles for different apico-basal tensions differences $\Delta\Gamma=\Gamma_\mathrm{a}-\Gamma_\mathrm{b}$, corresponding to different spontaneous curvatures.
 (d) Opening angles for different luminal volume constraints, where $V/V_\mathrm{VM}\approx1$ corresponds to the case without a luminal volume constraint. 
(e) Opening angles for rigidified shells different interfacial bending rigidities $\kappa_\mathrm{if}$, where $\kappa_\mathrm{if}=0$ is the regular BVM.
(f) Dependence on the basal apical tensions $\Gamma^{(5)}=\Gamma_\mathrm{a}^{(5)}=\Gamma_\mathrm{b}^{(5)}$ of pentagonal cells, with unchanged tensions in other cells.
(g) Dependence on basal (outer) tension $\Gamma_\mathrm{b}^{(5)}$ of pentagonal cells which is changed compared to the outer tension of the other cells. All other tensions are unchanged compared to the reference.
Markers are average angles and shaded regions display $5\%$ to $95\%$ quantiles. 
Single red lines in (f) and (g) are individual pentagonal cells and dotted vertical lines mark the reference case.}
\label{fig:modulation_pressure_rigidity}
\end{figure*}

Our experiments have revealed a coupling of tissue topology to cell shape and force generation.
In particular, the data hint at an active basal stabilization mechanism that controls
cellular shapes at topological defects (pentagons) in organoids. This agrees with a very recent
experimental report that basal contractility prevents extrusion in crypt regions \cite{Krueger_Clevers_Sciene_2025_Extrusion},  but it is challenging to
identify the exact mechanisms given the complex situation in organoids. In order to better understand 
the physical processes that might be involved in cell extrusion and shape stabilization, we turned back to the BVM.
Fig.~\ref{fig:modulation_pressure_rigidity} depicts the role of different tissue and cell 
properties on the opening angles at defects in simulated BVM-shells.
Fig~\ref{fig:modulation_pressure_rigidity}(a) displays exemplary shell shapes in modified BVM-shells.
For increasing shell size, we find that the effect of the topological defects on the opening angles decreases (Fig.~\ref{fig:modulation_pressure_rigidity}(b)).
In particular, for large shell sizes we see that some pentagonal cells also display negative opening angles, 
which is effected by cells bulging into the lumen as a consequence of the reduced background curvature.
Introducing spontaneous tissue curvature via an apico-basal tension difference $\Delta\Gamma=\Gamma_\mathrm{a}-\Gamma_\mathrm{b}$, modulates the opening angles slightly, 
but cannot fully stabilize cell shape (Fig.~\ref{fig:modulation_pressure_rigidity}(c)), 
even for a spontaneous curvature radius of hexagonal cells which corresponds to the shell radius, here $\Delta\Gamma\approx 0.14$.
In the following we will therefore usually neglect spontaneous curvature, i.e., consider $\Gamma\equiv\Gamma_\mathrm{a}=\Gamma_\mathrm{b}$.
Figure~\ref{fig:modulation_pressure_rigidity}(d) shows a shell whose luminal volume, $V$, was prescribed to be a multiple of the luminal volume of the relaxed VM shell, $V_\mathrm{VM}$, corresponding to an internal luminal pressure. 
Internal pressure is able to unfold pentagonal defect cells, 
cf.~Fig.~\ref{fig:modulation_pressure_rigidity}(a).
Contrary to the different opening angles for different number of neighbors, cf.~Fig.~\ref{fig:fig_shells_pentagonal_collapse}(c), increased pressure leads to convergence of the average opening angles of pentagons, hexagons and heptagons, see Fig.~\ref{fig:modulation_pressure_rigidity}(d). 
In addition, we also see that the distribution of opening angles in pressurized BVM-shells is less broad, more closely resembling the classical VM-case. This means that pressure can be used to suppress the buckling instability caused by the surface tensions. 
To estimate the experimental level of pressure-mediated stabilization, 
we calculated the theoretically expected aspect ratio \cite{Drozdowski_Boonekamp_2025_Force_inference, Drozdowski_PRR24_Topological_defects_VM_shells} 
of hexagonal cells in the limit of flat interfaces (the classical VM)
from inferred tensions (cf.~Fig.~\ref{fig:fig_exp_tens}(a); median relative tensions $\Gamma_\mathrm{a}\approx1.02$, $\Gamma_\mathrm{b}\approx1.06$, $\Gamma_\mathrm{l}\approx 0.97$) to be $\psi_\mathrm{theo}=h/A^{1/2}=2^{1/2}3^{-1/4} (\Gamma_\mathrm{a}+\Gamma_\mathrm{b})/\Gamma_\mathrm{l}\approx2.3$ with cell height $h$.
Experimentally we find $\psi_\mathrm{exp}=h^{3/2}V_\mathrm{cell}^{-1/2}\approx1.3$ yielding an experimental to theoretical shape ratio of $\psi_\mathrm{exp}/\psi_\mathrm{theo}\approx0.6$, which is consistent with a luminal pressure stretching the epithelium.
Comparing this ratio to the pressurized theoretical BVM case, we find similar values for a luminal volume of $V/V_\mathrm{VM}\approx2.25$ ($\psi_\mathrm{exp}/\psi_\mathrm{theo}=0.62)$, where cells have not fully unfolded. 
If we take average surface tensions of intestinal organoids from the literature, which are of the order $\langle\Gamma\rangle\approx 5\,\mathrm{nN}\,\mu\mathrm{m}^{-1}$ \cite{Yang_Cell21_Intestinal_organoid_VM},
we can roughly estimate the luminal pressure in the limit of flat interfaces (the classical VM) to be $P\approx 440\,\mathrm{Pa}$ (see Appendix~\ref{app:lumen_pressure} for details).
In our case, these pressures alone cannot fully 
explain smaller opening angles of pentagonal cells as observed experimentally.

To highlight the role of interface curvature, we added an interfacial bending energy to the BVM-minimization for each interface $\Omega_\mathrm{if}$
\begin{equation}\label{eq:interfacial_bending_energy}
    E_\mathrm{if} = \frac{\kappa_\mathrm{if}}{2} \int_{\Omega_\mathrm{if}} H_\mathrm{if}^2\, \mathrm{d}S,
\end{equation}
with interfacial bending rigidity $\kappa_\mathrm{if}$, and (total) mean curvature of the interface $H_\mathrm{if}$.
This model penalizes interfacial curvature energetically.
Fig.~\ref{fig:modulation_pressure_rigidity}(e) shows the opening angles, whose values quickly approach each other as $\kappa_\mathrm{if}$ is increased.
However, rigidifying the interfaces does not achieve as much convergence of the opening angles toward each other as luminal pressure.
In simulations we find the cells to have less curved interfaces, interpolating between the VM and the BVM, 
but this outcome does not agree with experimental observations, cf.~Fig.~\ref{fig:exp_mcol}.

Motivated by our experiments, we further changed the apical (inner) and 
basal (outer) tensions of the pentagonal defect cells to be different 
to the other cells to introduce topology-dependent variability (Fig.~\ref{fig:modulation_pressure_rigidity}(f)).
In this case, we see that increasing the defect tension 
leads to a reduction of pentagonal opening angles, 
where a transition through a shape consistent with hexagons toward negative opening angles occurs.
Note that reducing the tension can drastically increase the pentagonal opening angle, 
already hinting at a potential extrusion mechanism through a tension decrease.

Fig.~\ref{fig:modulation_pressure_rigidity}(g) shows
the effect of increased outer (basal) tensions at pentagonal cells, as found experimentally in the mCol-organoids. 
We also assumed a non-vanishing luminal pressure, as this is likely to reflect the experimental situation.
We find that the cells transition through a state consistent with hexagonal cells toward a state with approximately 
opposite opening angle, implying full cell shape control, cf. shell in Fig.~\ref{fig:modulation_pressure_rigidity}(a).
Note that the transition point is cell-specific, as it likely depends on the details of the neighborhood topology.
This suggests a potential feedback mechanisms between local topology (and induced stresses) and interfacial tensions.

Beyond surface tensions, we also considered an increase in
global basal line tensions, and found that 
it too can stabilize cell shapes, see Appendix~\ref{app:line_tensions}.
The basal stabilization mechanism observed in experiments thus proves to enable reliable control over cell shape.

\subsection{Shift of icosahedral buckling threshold}

The collapse of the luminal interface at pentagonal cells is reminiscent of the collapse observed earlier in the buckled icosahedral case in the VM for large apico-basal tensions $\Gamma$ and thus cell heights \cite{Drozdowski_PRR24_Topological_defects_VM_shells}.
In this case, a tissue-scale azimuthal stress from the pentagonal defect configuration is relaxed through conical buckling for large tissues, an effect described before for elastic shells in general \cite{Seung_88_PRL_Conical_instability_disclination}, leading to an icosahedral shape due to this effect occurring at the 12 pentagonal defects in the hexagonal tiling of the sphere.
This system-scale instability is well-known from viral capsids \cite{Lidmar_PRE_2003_icosahedral_instability_shells, Nguyen_PRE05_viral_capsid_spontaneous_curvature} and buckyballs \cite{Witten_EuPhysLett93_Fullerene_Ball_icosahedral_instability}.
On a single-cell scale, at these topological defects luminal interfaces can collapse due to the large necessary cellular opening angle at the icosahedral tips, quite similarly to what we observe in the BVM.

\begin{figure}[!t]
	\centering
	\includegraphics[width=\columnwidth]{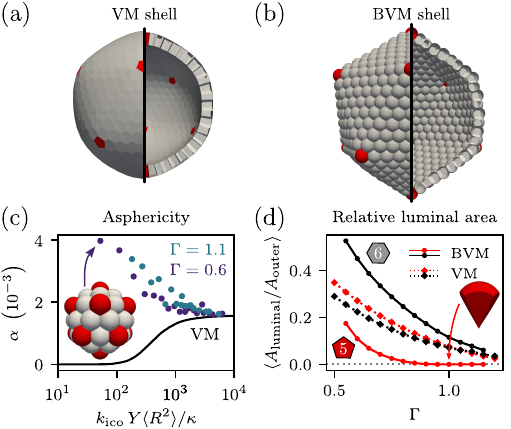}
	\caption{Lowered buckling threshold for icosahedral instability in the bubbly vertex model (BVM) compared to the vertex model (VM).
 (a,b) Icosahedral shells with Caspar Klug indices $(8,0)$ with tension $\Gamma=0.6$ in the VM (a) and the BVM (b).
 (c) Asphericity $\alpha$ in the BVM-shell (symbols) and of the continuum curve describing the VM (taken from Refs.~\cite{Lidmar_PRE_2003_icosahedral_instability_shells, Drozdowski_PRR24_Topological_defects_VM_shells}). Inset shows shell with size $(2,0)$, $\Gamma=0.6$. 
 Radii are rescaled by the F\"oppl-von K\'arm\'an number and with a correction $k_\mathrm{ico}$ from non-linear elasticity.
 (d) For a small shell (Caspar Klug indices $(2,0)$) the average ratio of luminal and outer interfacial area in the BVM (solid) and VM (dashed). Inset shows pentagonal cell with collapsed luminal interface ($\Gamma=1.0$). Pentagonal cells are shown in red.
 }
\label{fig:fig_ico_shells_bubbly}
\end{figure}

To be able to compare with these results for the VM, 
we now also consider spherical shells with icosahedral symmetry, following the Casper-Klug construction \cite{Caspar_Klug_1962_Physical_principles_construction_regular_viruses}.
The Caspar-Klug indices $(T_1,T_2)$ 
describe the steps in the underlying hexagonal lattice to go from one pentagonal defect cell to another with respect to lattice vectors with an opening angle of $\pi/3$.
If we compare the results for the VM (Fig.~\ref{fig:fig_ico_shells_bubbly}~(a))
versus the results for the BVM (Fig.~\ref{fig:fig_ico_shells_bubbly}~(b)),
we see that the buckling instability in the BVM
is much stronger, for the same number of cells.
It also occurs at smaller radii than the buckling
instability in the VM, which only occurs for
large system size. To quantify this effect,
we measured the asphericity
\begin{equation}
\alpha = \frac{\langle (R-\langle R\rangle)^2\rangle}{\langle R^2 \rangle},
\end{equation} 
where $R$ is the radial distance 
of a cell (determined as the average 
of the centers of mass of the apical and basal interfaces)
from the shell center and $\langle \cdot \rangle$ is the multicellular average.
For the VM, this quantity has been shown before to follow a 
continuous curve as a function of the 
rescaled shell size, corresponding to the F\"oppl-von K\'arm\'an number
$\gamma_\mathrm{FvK}\equiv Y_\mathrm{eff}R^2/\kappa$, with bending rigidity $\kappa$, radius $R$, and (effective) Young's modulus $Y_\mathrm{eff}$, corrected here for effects from nonlinear elasticity with a numerical
factor $k_\mathrm{ico}$ \cite{Drozdowski_PRR24_Topological_defects_VM_shells}.
As shown in Fig.~\ref{fig:fig_ico_shells_bubbly}(c), the asphericity is much larger for small shells in the BVM compared to the VM. Details on the axis scaling are given in Appendix~\ref{app:continuum_params}.

We also find that the asphericities in the BVM do not fall on one curve, but rather deviate from an average curve, which results from both the broad distributions of hexagonal opening angles described above and therefore less regular structures, and from cellular positioning for small cell numbers on the icosahedral surface due to undersampling of the icosahedral surface.
Similarly to the VM, we have a collapse of the luminal interface for larger $\Gamma$, which is much more pronounced in the BVM with a stronger deviation of the luminal-to-outer area ratios between pentagonal and hexagonal cells for small spheres, cf.~Fig.~\ref{fig:fig_ico_shells_bubbly}~(d).
Even shells as small as $(2,0)$, i.e. only one hexagonal cell between the pentagonal cells, show a big
difference between pentagonal and hexagonal cells, and thus adopt a strongly icosahedral shape.
We thus see that the buckling threshold is reduced to effectively one neighboring cell in the BVM, rendering the icosahedral tissue-scale instability much more relevant in the case of curved cellular interfaces in the BVM.

\section{Continuum theory for extrusion of a defect cell in a mean-field tissue}\label{sec:continuum}

\subsection{Parameterizing the defect cell}

\begin{figure}[!t]
	\centering
	\includegraphics[width=\columnwidth]{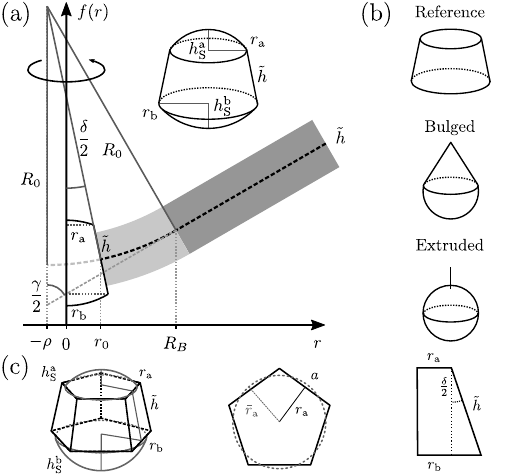}
	\caption{Schematic depiction of the defect cell in a mean field tissue. 
	(a) The tissue profile is shown as a radial function in cylindrical coordinates. At the center is the defect cell, which consists of a base of a cone with spherical caps on the apical and basal sides. This defect is surrounded by mean-field tissue that forms a spherical cap segment (light grey) and for larger distance a cone (dark grey). 
	(b) The model can describe different defect cell geometries, namely, the vertex-model reference case ($h_\mathrm{S}^\mathrm{a}=h_\mathrm{S}^\mathrm{b}=0$), a completely bulged case ($r_\mathrm{a}=0$, $\delta/2 >0$) and a fully extruded case ($r_\mathrm{a}=0$, $\delta/2=0$).
	(c) For pentagonal cells we consider a geometrical correction, where we assume a pentagonal pyramid base with two spherical caps. The radii of the caps, $\bar{r}_\mathrm{a,b}$, are chosen such that the area of the caps' disc bases, the circles, match the pentagonal areas (center). and the slant height $\tilde{h}$ is considered at the lateral faces' centers.}
\label{fig:extrusion_schematic_defect_cell_in_mean_field_tissue}
\end{figure}

To understand the nature of the significant lowering of the icosahedral buckling threshold for the BVM compared to the VM 
and expand the model to capture cell extrusion, 
we now investigate the BVM analtically in a continuum setting.
We focus on the icosahedral tip and the conical instability to first develop the continuum model in a setting amenable to comparison to numerical simulations.
At the site of the disclination defect we excise the continuous material and directly consider the energy of the pentagonal cell, following the description of the BVM, cf.~Eq.~(\ref{eq:continuum_bubbly_vertex_energy}). 
Fig.~\ref{fig:extrusion_schematic_defect_cell_in_mean_field_tissue}(a) depicts the geometry we assume. 

At the center of the disc of material is the defect cell, which we consider first.
We assume rotational symmetry and thus the cutout of the tissue is the base of a cone.
The cell is allowed to bulge outward via interfacial curvature, 
which is captured by considering spherical caps at the apical and basal sides of the defect cell.

With the chosen parameterization we are able to describe different defect cell morphologies, 
i.e. a vertex model case with flat apical and basal interfaces, 
a bulged case with a collapsed interface, 
and an extruded case (Fig.~\ref{fig:extrusion_schematic_defect_cell_in_mean_field_tissue}(b)).
Due to the pentagonal nature of the defect, we modify the conical shape to reflect the pentagonal structure by considering a pentagonal pyramid base, see Fig.~\ref{fig:extrusion_schematic_defect_cell_in_mean_field_tissue}(c).
For this we introduce a geometrical factor $\zeta_\mathrm{geom}$, which relates the basal and apical polygonal edge lengths, $b$ and $a$, respectively, to the areas such that for the apical area $A_\mathrm{a} = \zeta_\mathrm{geom} a^2$ and similarly for the basal side.
Additionally, we introduce $\eta_\mathrm{geom}$, which relates the radius of a regular polygon's incircle $r_\mathrm{a,b}$, i.e. the largest circle that fits into the regular polygon, to the edge length, e.g., for the apical side $r_\mathrm{a}=\eta_\mathrm{geom} a$. 

For a bulged pentagonal cell we consider the pyramid base with apical/basal spherical caps. 
We assume the caps to have the same disc area as the apical/basal pentagons, neglecting the slight mismatch of the two different shapes, cf.~Fig.~\ref{fig:extrusion_schematic_defect_cell_in_mean_field_tissue}(c).
This is consistent with simulated cell shapes in the BVM as shown in Fig.~\ref{fig:fig_shells_pentagonal_collapse}(a).
For the spherical caps we denote the base disc radii with $t_\mathrm{a}$; the cap's base area is $\pi t_\mathrm{a}^2$;
this implies $\pi t_\mathrm{a}^2 = \zeta_\mathrm{geom} a^2$ for the apical side and analogously for the basal side.
The radius of the apical spherical cap's base then relates to the edge length via $t_\mathrm{a}=\sqrt{(\zeta_\mathrm{geom}/\pi)}\, a$ (analogously for basal).

Henceforth we denote the side to which the tissue buckles (positive $z$) the 
apical side and the opposite side the basal one, as this corresponds to the situation in the experimental mCol-organoids when considering closed shells.
Note that the sign of curvature at the villus tip in the small intestine is opposite.
The opening angle of the pyramid at the lateral face centers, $\delta/2$, the apical inradius $r_\mathrm{a}$ and the tissue height $\tilde{h}$ define the base part.
To consider the energy change due to extrusion and bulging, we define as the reference case the vertex model case, where the defect cell is not bulged and induces the same curvature as the tissue would. 
The spherical caps of the defect cell are uniquely defined by the basal and apical radii and the height of the caps ${h_\mathrm{S}^\mathrm{a,b}}$ (cf.~Fig.~\ref{fig:extrusion_schematic_defect_cell_in_mean_field_tissue}).

The radius of curvature of the spherical caps $R_\mathrm{S}^\mathrm{a,b}$ depends on the pressure differences between the cell and the surrounding medium $\Delta P_\mathrm{a,b}$ at the corresponding side via the Young-Laplace law
\begin{equation}
    \Delta P_\mathrm{a,b} = \frac{2 \Gamma_{a,b}}{R_\mathrm{S}^\mathrm{a,b}}.
\end{equation}
If we assume equal pressure differences, i.e., no pressure difference between the basal and apical medium, the radii are related via
\begin{equation}
    \frac{R_S^a}{\Gamma_\mathrm{a}} = \frac{R_S^b}{\Gamma_\mathrm{b}}.
\end{equation}
As described earlier, we assume $\Gamma_\mathrm{a}=\Gamma_\mathrm{b}$ in the analytical treatment, i.e., no apico-basal polarity.
Using this relation, we express the basal cap height via the apical one, assuming that extrusion only occurs on the 
outer (basal) side (similar to the sign of curvature at the intestinal villus tip) and therefore choosing the smaller of two possible cap heights.

The volume of the cell then implicitly defines $h_S^a$ via 
\begin{multline}
    1 = \frac{\tilde{h}\cos(\delta/2)}{3} \zeta_\mathrm{geom} \left(a^2 + ab + b^2\right) \\
    + \frac{h_\mathrm{S}^\mathrm{b}}{6} \left( 3 \zeta_\mathrm{geom}b^2 + \pi {h_\mathrm{S}^\mathrm{b}}^2\right) + \frac{h_\mathrm{S}^\mathrm{a}}{6} \left( 3 \zeta_\mathrm{geom}a^2 + \pi {h_\mathrm{S}^\mathrm{a}}^2\right),
\end{multline}
with the lateral (slant fraction) height $\tilde{h}$. 
The last two terms correspond to the spherical caps and vanish for the non-bulged reference cell as $h_\mathrm{S}^\mathrm{b}=h_\mathrm{S}^\mathrm{a}=0$. 
The apical and basal sides are related to each other via the opening angle, $b=a+\sin(\delta/2)\tilde{h}/\eta_\mathrm{geom}$, cf.~Fig.~\ref{fig:extrusion_schematic_defect_cell_in_mean_field_tissue}(c). 

In a continuum approach we would not resolve the defect cell and the surrounding spherical cap with spherical radius $R_0$ would determine the opening angle, i.e., $\sin(\delta/2) = \eta_\mathrm{geom} \left( 1+ \frac{R_0+\tilde{h}/2}{R_0 - \tilde{h}/2}\right) a/2R_0$.
However, resolving the defect cell explicitly allows us to consider a vertex model later, which goes beyond this approximation, even if we do not allow for interfacial curvature.

With known cap heights we can write down the energy of the bubbly defect cell by adding the lateral surface energies and the surface energies of the spherical caps to obtain, where the cell is given by a regular $n$-gon,
\begin{multline}\label{eq:extrusion_continuum_bubbly_core_energy}
    E_\mathrm{core}^\mathrm{bulged} = \Gamma_\mathrm{a} \left( \zeta_\mathrm{geom} a^2 + \pi {{h_\mathrm{S}^\mathrm{a}}}^2\right) + \Gamma_\mathrm{b} \left( \zeta_\mathrm{geom} b^2 + \pi {{h_\mathrm{S}^\mathrm{b}}}^2\right) \\
    + \frac{n\tilde{h}}{4} \left(2a + \frac{\tilde{h}}{\eta_\mathrm{geom}} \sin(\delta/2)\right).
\end{multline}
For a pentagonal defect we have 
$n=5$,
$\zeta_\mathrm{geom}=\sqrt{5(5+2\sqrt{5})}/4$, and $\eta_\mathrm{geom}=1/\left(2\sqrt{5-\sqrt{20}}\right)$.
We assume that the lateral slant height of the pyramid base is the equilibrium height of the surrounding tissue $\tilde{h}$ and thus the only two free parameters are $a$ and $\delta$. 
To describe the full tissue we additionally need the background curvature $R_0$, which relates to the buckling radius $R_\mathrm{B}$ via the conical opening angle $\gamma$, cf.~Fig.~\ref{fig:extrusion_schematic_defect_cell_in_mean_field_tissue}(a).

\begin{figure}[!t]
	\centering
	\includegraphics[width=\columnwidth]{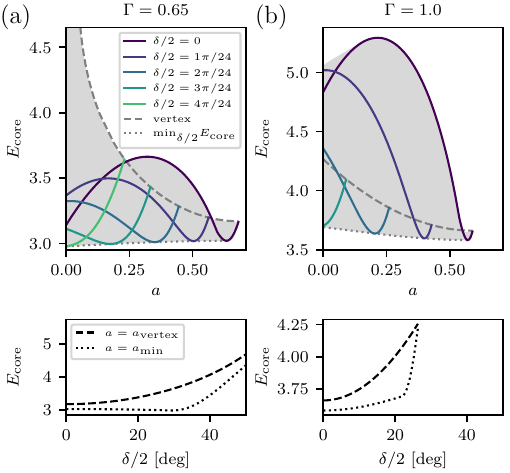}
	\caption{The core energy for the different defect cell configurations for (a) small apical/basal surface tension $\Gamma$ and (b) large $\Gamma$. The solid colored curves in the upper panels are the energies as a function of apical edge length $a$ for fixed opening angles $\delta/2$. The curves end at the vertex line (dashed) at which $a$ is maximal for the angle as such that the apical and basal faces do not bulge.  All possible states and energies which can be reached with $\delta/2$ and $a$ such that the volume constraint is fulfilled are shaded in grey.
The minimum line (dotted) indicates the minimal energy for given $a$. In the lower panels the vertex line and the minimum line are parameterized with $\delta/2$. The minimum line is obtained by finding the edge length $a_\mathrm{min}$ that minimizes $E_\mathrm{core}$ for a given $\delta/2$. We consider a pentagonal defect with equilibrium tissue height of the hexagonal vertex model, i.e., $\tilde{h}=2^{1/3}\ 3^{-1/6}\ (2\Gamma)^{2/3}$.}
\label{fig:extrusion_core_energy}
\end{figure}

Fig.~\ref{fig:extrusion_core_energy} shows the dimensionless core energy, Eq.~(\ref{eq:extrusion_continuum_bubbly_core_energy}),
for different opening angles $\delta/2$ as a function of the apical edge length $a$ for (a) $\Gamma=\Gamma_\mathrm{a}=\Gamma_\mathrm{b}=0.65$ and (b) $1.0$, respectively. 
Edge length $a$ is bounded from above 
by the volume constraint in the unbulged vertex model case with $h_\mathrm{S}^\mathrm{b}=h_\mathrm{S}^\mathrm{a}=0$. 
For each angle the energy initially decreases 
if we decrease $a$ with concomitant outward bulging (top panel), as the area to volume ratio decreases.
For even smaller $a$ the energy usually increases again, suggesting that a configuration with curved interfaces is always energetically more favorable.

Increasing the angle leads to an increase in the core energy in the vertex model case (cf.~dashed line at which the curves end in the top panel) and shifts the minimum of the energy toward smaller $a$ if we allow for bulging. 
If we compare the vertex model energy with the minimal energy in the bubbly vertex model for given opening angles (and thus surrounding tissue curvature), we find a smaller energy in the latter case, see lower panel. 

Strikingly, as we increase $\delta/2$ and thus bend the defect cell, we either obtain a smaller increase of energy in the bubbly case compared to the vertex model ($\Gamma=1.0$) or even an energy release ($\Gamma=0.65$).
As this is related to the bending rigidity, this drastically changes the behavior of the overall tissue. 
The potential to bulge outward therefore decreases the necessary bending energy in the defect core and can even lead to an energy release. 
This energy can then compensate for energy contributions in the surrounding tissue, allowing for a different equilibrium shape than in the two-dimensional elastic sheet.
This effect explains the observed concentration of Gaussian curvature in the defect cell, as it is energetically cheaper to go into the buckled state for an icosahedral shell.

\subsection{Deformation of the mean-field tissue}

For the surrounding mean field tissue we assume that the inner radius of the inner tissue segment (cf. light grey part in Fig.~\ref{fig:extrusion_schematic_defect_cell_in_mean_field_tissue}(a)) corresponds to the polygonal inradius of the defect cell, 
neglecting mismatch contributions here stemming from the shape difference between the cellular pentagonal mid-plane and circular cut-out.
This inner cap segment is assumed to follow a circular arc with radius of curvature $R_0$ in cylindrical coordinates, cf.~Fig.~\ref{fig:extrusion_schematic_defect_cell_in_mean_field_tissue}(a).
An outer segment (cf. dark grey part) is assumed to be conical, in agreement to the result in moderately bent plates with a disclination defect.

We use the description as a moderately bent plate as derived in Ref.~\cite{Drozdowski_PRR24_Topological_defects_VM_shells}.
The energy change due to the defect deformation can then be split into different contributions. 
First, the bending energy will change, both for the mean and Gaussian curvature contributions, due to changes in the curvature radii and opening angles in the inner cap segment. 
Second, the stretching energy will change as the azimuthal strain stemming from the pentagonal defect can be relaxed through changes in the curvature. 
For simplicity we will neglect radial stretching contributions at the interfaces by considering stress free boundary conditions at the contact points of defect cell, 
cap segment and outer cone segment and at the outer edge of the cone. This is justified numerically in Appendix~\ref{app:mean_field_energies}.

Introducing the bending energy density $e_\mathrm{bending}$ and the stretching energy density $e_\mathrm{stretch}$, the total energy of the tissue $\Omega$ then reads
\begin{equation}\label{eq:total_energy_bending_stretching}
    E = \int_\Omega \left( e_\mathrm{bending} + e_\mathrm{stretch}\, \right) \mathrm{d}S.
\end{equation}

For the spherical topology of the tissue which we consider here, we consider the conical deformation of a disc with the pentagonal defect in the center,
as this has been shown to capture the icosahedral instability of such a regular sphere \cite{Drozdowski_PRR24_Topological_defects_VM_shells, Lidmar_PRE_2003_icosahedral_instability_shells}.

The different energy contributions for bending and stretching for this mean-field tissue parameterization are derived 
in Appendix~\ref{app:mean_field_energies}. 
This parameterization with a mean-field model has been verified on icosahedral vertex model 
shells without interfacial curvatures,
where the mentioned decomposition into a defect cell and a mean-field tissue is able to describe the tip much better than an approach purely based on continuum theory, cf. Appendix~\ref{app:verification_mean_field_ico_shells}.

\subsection{Extrusion of a bubbly defect cell in a mean-field vertex model tissue at icosahedral tips}
We now consider a bubbly defect cell in a mean-field tissue of icosahedral symmetry, i.e., $h_\mathrm{S}^\mathrm{b}$ and $h_\mathrm{S}^\mathrm{a}$ may deviate from zero.

In the following we assume the mean-field tissue parameters to be from the classical (non-bubbly) VM, thus only allowing for bulging of the defect cell. 
In principle, however, mean-field parameters could be derived for the BVM  in a similar fashion as for the  VM, or could be taken from experimental measurements for the tissue properties.

\begin{figure*}[!ht]
	\centering
	\includegraphics[width=\textwidth]{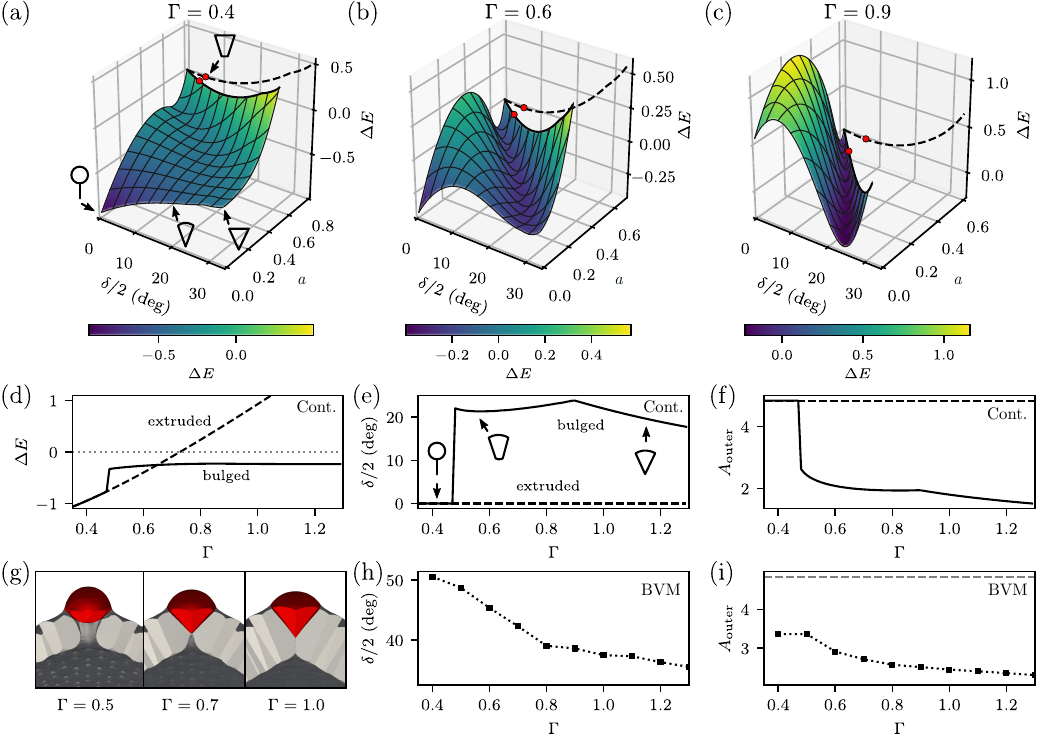}
\caption{Energy landscapes in configuration space and the resulting defect cell shapes.
(a-c) Energy landscapes with coordinates $(\delta/2, b)$ for different surface tensions $\Gamma$. The black line marks the vertex model limit without bulging of the defect cell and the red circle marks the minimum in this limit. The dashed line is the projection onto the $(\Delta E, \delta/2)$ plane of this limit. The color codes the energy difference $\Delta E$ with respect to the reference energy, which is the vertex model minimum. The corners of the landscape at $(\delta/2,b)=(0, 0)$ and $(\delta_\mathrm{max}/2, 0)$ correspond to the extruded and maximally bent situations,
respectively, cf. Fig.~\ref{fig:extrusion_schematic_defect_cell_in_mean_field_tissue}.
(d) Continuum model energy differences from the VM reference $\Delta E$ of the minimal energy configuration (solid), starting from the VM reference, and of the extruded configuration (dashed).
(e,f) Corresponding defect opening angles $\delta/2$ and outer interfacial area $A_\mathrm{outer}$.
(g) For the BVM with interfacial bending rigidity the defect cell is extruded for small apico-basal tension $\Gamma=0.5$, and partially extruded for large tensions $\Gamma=0.7$ and $\Gamma=1.0$.
(h,i) Corresponding opening angles and outer interfacial area for the BVM.
Shells of Caspar-Klug size $(T_1,T_2)=(12,0)$ with bending rigidity $\kappa_\mathrm{if}=5$ in dimensionless units.}
\label{fig:energy_landscapes_extrusion}
\end{figure*}

The total energy is the sum of the contributions from the defect, the cap and the outer cone region.
We consider the energy difference to a reference state to quantify the energy release due to bulging by considering the energy difference $\Delta E$ from the vertex model case with minimal energy.
The latter corresponds to the full defect model described in Appendix~\ref{app:verification_mean_field_ico_shells}, 
see~Fig.~\ref{fig:extrusion_comparison_mean_field_VM}.

We find that for all configurations characterized by $\delta/2$ and $a$ the energy has a unique minimum in the tissue buckling radius $R_\mathrm{B}$. 
We assume this radius to be minimal for every defect cell shape 
and thus can describe all possible shapes by the variables $(\delta/2, a)$, which constitutes the configuration space.
Both parameters are bounded from below by $0$, where $(0,0)$ is possible and corresponds to the fully extruded case, 
and from above by the volume constraint of the defect cell.

Fig.~\ref{fig:energy_landscapes_extrusion}(a,b,c) show energy landscapes in configuration space for different surface tensions $\Gamma$.
For now we assume identical cell properties, i.e. the tension in the defect cell matches the one of the surrounding tissue.
We notice that, by bulging outward at the defect, the energy of the tissue can be reduced considerably
in comparison to the classical vertex model reference case.
For small $\Gamma$ more energy can be released by bulging outward than for larger $\Gamma$. 
For $\Gamma>0.5$ we observe that two minima form in the energy, one in the extruded case $(\delta/2,a)=(0,0)$,
and one in a bulged case with very small $a$.
The energy minimum in extrusion is higher than the bulged one for larger $\Gamma$ 
and an energy barrier has to be overcome from both the reference and the global minimum configuration to extrude the defect.

This effect is to be expected, 
because we have assumed that the bulged membrane has surface tension $\Gamma$ and thus the fully extruded cell (surrounded by said membrane) 
will have a higher energy due to this larger surface tension.
For $\Gamma<0.5$ on the other hand we find that the extruded case is minimal in energy and can directly be reached without passing over an energy barrier. 
The case $\Gamma=0.5$ is special because it corresponds to cells with aspect ratio $\approx 1$, 
as the lateral energies only contribute half to the total energy, cf.~Eq.~(\ref{eq:continuum_bubbly_vertex_energy}), and thus energy costs for lateral and apico-basal interfaces do not differ.
For $\Gamma<0.5$ the lateral membrane is effectively more expensive than the apical membrane.

We now numerically determined the (local) minimal energy configurations for different $\Gamma$, starting from the VM reference state.
Fig.~\ref{fig:energy_landscapes_extrusion}(d,e,f) depict the resulting minimal energy configurations.
Above a critical tension $\Gamma$ the luminal (basal) interface collapses but the angle remains finite, indicating a completely bulged state.
The fully extruded state at $(\delta/2,a)=(0,0)$ always represents a possible configuration, but its energy increases with $\Gamma$.
Looking at the energy difference from the VM state, a configuration with interfacial curvature is always energetically more favorable, cf.~Fig.~\ref{fig:energy_landscapes_extrusion}(e,f).
This is not surprising as the surface area can be minimized by the cell becoming more spherical.
The extruded case, on the other hand, requires more energy for large apico-basal tension, as we need to wrap the cell in the energetically (relatively) expensive apical interface. 
Lowering the tension allows for an energy release in the extruded case, which can become lower than in the bulged state.
Note, however, that a bulged state persists beyond the point at which the extruded case is energetically more favorable 
and that an energy barrier exists in configuration space, which prevents the cell from being directly extruded from the reference VM configuration, cf.~Fig.~\ref{fig:energy_landscapes_extrusion}(b).

Considering icosahedral shells in the BVM and adding a bending rigidity across triangulation points of hexagonal cells' interfaces enables a numerical simulation similar to our continuum approach of a bubbly defect cell surrounded by VM cells, 
cf. Fig.~\ref{fig:modulation_pressure_rigidity}(g).
We thus consider an additional bending energy, Eq.~(\ref{eq:interfacial_bending_energy}).

Fig.~\ref{fig:energy_landscapes_extrusion}(g) shows configurations of the defect cell in this BVM with bending rigidity for different apicobasal tensions $\Gamma$.
For $\Gamma\leq0.5$ we see that indeed the pentagonal cell is pushed out of the monolayer, only being attached because the lateral interfaces cannot be broken in the simulations. 
This is also visible in the the parameters quantifying the configurations in Fig.~\ref{fig:energy_landscapes_extrusion}(h,i) 
The luminal interfaces bend outward despite the energetic penalty to facilitate this extrusion process.

For intermediate $\Gamma$ the neighbor cells are deformed due to finite $\kappa_\mathrm{if}$, while
for large $\Gamma$ a completely bulged state is found, as predicted by the continuum model.

\subsection{Extrusion at pentagonal and hexagonal defects in spherical shells with curvature screening}
\begin{figure*}[!ht]
	\centering
	\includegraphics[width=\textwidth]{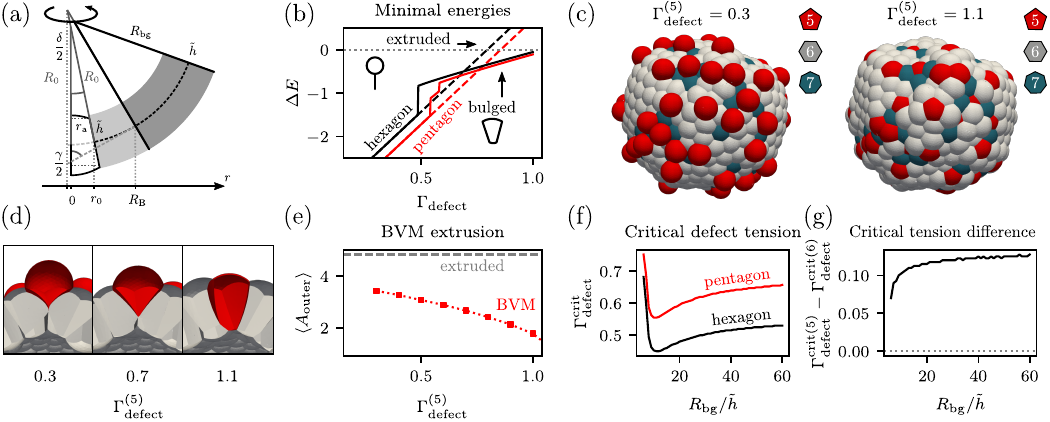}
\caption{Extrusion through lower defect tension in hexagons and pentagons.
 (a) Schematic depiction of defect in mean-field tissue with background curvature (radius $R_\mathrm{bg}$).
 (b) The energy difference for continuum model configurations with different defect tensions $\Gamma_\mathrm{defect}$ from the VM case with identical defect tension $\Gamma_\mathrm{defect}=\Gamma$. Solid line is local minimization from VM case, dashed line is extruded case. Red is pentagonal, black hexagonal defect cell.
 (c) BVM-shells from Fig.~\ref{fig:fig_shells_pentagonal_collapse} in which pentagonal cells have different apico-basal tensions $\Gamma_\mathrm{defect}^{(5)}$, displaying partially extruded morphologies that depend on the tension. Pentagons in red, heptagons in blue.
 (d) Cross section of a pentagonal cell surrounded by hexagonal cells being extruded upon lowering of $\Gamma_\mathrm{defect}^{(5)}$ in the BVM.
 (e) Average outer interfacial area of BVM-shells; horizontal dashed line corresponds to fully extruded cells.
 (f) Change of the critical defect tension $\Gamma_\mathrm{defect}^\mathrm{crit}$ in the continuum model. at which the bulged state becomes unstable toward extrusion for different background radii $R_\mathrm{bg}$.
 (g) Difference of critical tensions in pentagonal and hexagonal cells in (f).
 Parameters: $\Gamma=0.9$; (b,f) $R_\mathrm{bg}/\tilde{h}=5$.}
\label{fig:critical_tension_defect}
\end{figure*}
We generalize our continuum description to spherical shells with curvature screening, as we have also seen an effect of topological defects in less structured monolayers, cf.~Fig.~\ref{fig:fig_shells_pentagonal_collapse}.
For this we assume the topological defect to only induce azimuthal strain locally around the defect.
In simulations we see that pentagons with one neighboring ring of hexagonal cells exist in spherical shells due to the necessary excess topological charge (12 more pentagons than heptagons), cf.~Fig.~\ref{fig:fig_shells_pentagonal_collapse}.
Thus, we consider a defect cell, surrounded by a cap region with varying radius of curvature but fixed size (i.e. fixed buckling radius $R_\mathrm{B}$), which in turn is surrounded by background tissue as depicted in Fig.~\ref{fig:critical_tension_defect}(a). 
The background tissue is assumed to be mechanically relaxed and not considered in the following:
the curvature is screened accordingly that a spherical configuration is obtained, cf.~Fig.~\ref{fig:fig_shells_pentagonal_collapse}(b), and topological defects at most affect their neighbors.
The in-radius for a hexagonal cell is
\begin{equation}
    r_0^\mathrm{hex} = \sqrt{\frac{1}{\tilde{h}\zeta_\mathrm{geom}^\mathrm{tissue}}}\eta_\mathrm{geom}^\mathrm{tissue} = 3^{-1/6} 2^{-2/3} \left(2\Gamma\right)^{-1/3}.
\end{equation}
We now assume for the radius of the cap along the radial axis $R_\mathrm{B}=3r_0^\mathrm{hex}$, i.e. one neighboring ring of cells.
The background curvature radius $R_\mathrm{bg}$ only enters via the opening angle (similarly to the conical opening angle), i.e.
\begin{equation}
    \pi/2 - \gamma/2 = \arcsin\left(R_\mathrm{B}/R_\mathrm{bg}\right).
\end{equation}

For this configuration the total energy is now identical to the contributions described in Appendix~\ref{app:verification_mean_field_ico_shells} 
(cf.~Eq.~(\ref{eq:extrusion_sum_all_energies_cone})), but without the cone energy and with prescribed $R_\mathrm{B}$ and new angle $\gamma/2$, i.e.
\begin{equation}
\begin{aligned}
    E_\mathrm{tot} &= E_\mathrm{core} + E^\mathrm{mean}_\mathrm{bending} + E^\mathrm{Gauss}_\mathrm{bending} \\
    &\quad + E_\mathrm{stretch} + E_\mathrm{bndry}.
\end{aligned}
\end{equation}

We now also want to consider extrusion of a hexagonal cell, for which we have no azimuthal stretching 
(a disclinicity of $s=0$ in Eq.~(\ref{eq:extrusion_delta_s_with_pentagon}) in Appendix~\ref{app:mean_field_energies}) and consider $n=6$ lateral faces.
The geometric constants then read $\eta_\mathrm{geom}=3^{1/2}/2$ and $\zeta_\mathrm{geom}=3^{3/2}/2$.
We have found that lower apico-basal tensions lead to extrusion, because the cell has to be completely engulfed by the apical interface in the extruded case.
This suggests that one potential mechanism of extrusion is the lowering of apico-basal tension in the defect cell.
Fig.~\ref{fig:critical_tension_defect}(b) shows the energy differences from the VM reference for different defect tensions $\Gamma_\mathrm{defect}$ in a mean field tissue of tension $\Gamma=0.9$.
We see that pentagonal energies are generally smaller in deformed states due to the relaxation of azimuthal strain when localizing Gaussian curvature in the defect.

For both hexagons and pentagons we see that the extruded and bulged states become energetically more favorable as $\Gamma_\mathrm{defect}$ decreases. 
Below a critical tension $\Gamma_\mathrm{defect}^\mathrm{crit}$ the bulged state becomes unstable toward the extruded state.

In BVM-simulations, we find strong outward bulging reminiscent of intermediate states in cell extrusion upon lowering of the apical and basal defect tension, without achieving full extrusion due to the inability of closing the forming hole in the simulations (Fig.~\ref{fig:critical_tension_defect}(c,d,e)).
In Fig.~\ref{fig:critical_tension_defect}(f) these critical tensions for pentagons and hexagons are plotted as functions of the background curvature radius $R_\mathrm{bg}/\tilde{h}$.
For a small radius the critical tension is high, implying that extrusion is easy. 
The reason for this is that the volume constraint does not allow for a full relaxation of the Gaussian curvature in the defect, because we have assumed that $\delta/2<\pi/2-\gamma/2$, i.e. the cap region cannot accommodate negative curvature.
This leads to a less stable bulged state and thus earlier instability toward extrusion.
For intermediate background radii the bulged state is able to stabilize by relaxing the remaining Gaussian curvature in the defect.
Increasing the background radius further then leads to less stability of the bulged state, because the possible opening angle is small anyways.

In the pentagonal cell we see an intermediate state in Fig.~\ref{fig:critical_tension_defect}(b):
for intermediate tensions a local minimum of full bulging ($a=0$, $\delta/2>0$) forms, which then becomes unstable toward the extruded case for lower tensions (cf.~Fig.~\ref{fig:energy_landscapes_extrusion}(a,b)).
In general we see that for decreasing defect tensions in both pentagons and hexagons the energy landscape loses local minima corresponding to bulged states, as the extruded case becomes increasingly more energetically favorable.

Fig.~\ref{fig:critical_tension_defect}(g) depicts the difference in the critical defect tensions of pentagons and hexagons, $\Gamma_\mathrm{defect}^{\mathrm{crit(5)}}-\Gamma_\mathrm{defect}^{\mathrm{crit(6)}}$.
The necessary tension decrease for overcoming the energy barrier for extrusion in hexagons is approximately $10\%$ of the total tension higher than for pentagonal cells.
While this suggests pentagonal cells to be more prone to extrusion, it also does not constitute a difference large enough for other cellular processes to not be able to overcome the barrier for hexagons as well.
Introducing additional luminal line tensions yields a destabilization of a bulged state with a finite size luminal interface, does not contribute, however, to the extrusion process without an involvement of lateral tensions, see Appendix~\ref{app:line_tensions}.

\section{Discussion}\label{sec:discussion}

Here we have introduced the bubbly vertex model (BVM) for multicellular epithelial monolayers as a generalization of the standard vertex model (VM). In contrast to the VM,
the BVM allows for curved interfaces, which are
known to occur in flat epithelia and even more so
in curved ones. Because curved epithelial sheets
naturally have topological defects due to 
Euler's polyhedron theorem, we asked
if cell bulging and extrusion might occur naturally
at such defects. By comparing the BVM with the VM,
we found that allowing for cell interface curvature 
dramatically lowers the conical buckling threshold around such defects and yields a bulging instability.

To explore the significance of these theoretical predictions, we have performed
experiments with spherical mouse colon organoids, which are
an appropriate model system for the spherical epithelia described
by our theory. Because extrusions are
hard to capture due to their speed, we focused on cell shape
at topological defects. We found that indeed cell shape is
different at pentagonal cells, but differently from the predictions
of the BVM, the opening angles were decreased rather than increased.
This suggests the existence of some mechanisms for stabilization.
Very recently, basal contractility has been reported as
such a mechanism \cite{Krueger_Clevers_Sciene_2025_Extrusion}.
Leveraging a force inference pipeline that we have developed earlier \cite{Drozdowski_Boonekamp_2025_Force_inference}, we indeed found an increased basal interfacial tension.
Perturbation of the actomyosin cytoskeleton increased the bulging, 
showing its involvement in actively stabilizing cell shape.
We found pentagonal mid-plane areas to be smaller and negatively correlated with basal tensions, potentially facilitating stabilization via mechanosensing.
In simulations we found this strategy of cell-specific tension variability to yield full control over cell shapes.
We also showed that tissue-scale forces must be at play, in particular
pressure, which we estimate to be of the order of $440\,\mathrm{Pa}$.
This is one order of magnitude smaller than reported for 
epithelia compressing hydrogel spheres \cite{Wu_PNAS25_epithelial_homeostasis_pressure},
but likely larger than in mouse small intestine organoids, which
use lower pressures as a mechanism for budding \cite{Yang_Cell21_Intestinal_organoid_VM}.

In order to better understand the physical mechanism
behind the instability, we developed a theory 
of a discrete defect cell in a mean-field tissue.
This approach provides a mechanistic model for extrusion-like cellular shapes and it suggests that lowering the apico-basal tension in the cell to be extruded, 
and thus lowering the contractile forces and the mechanical stability of its cortex, provides a mechanism to extrude cells. 
This is consistent with recent experiments, showing extrusion upon cortical ablation in intestinal organoids 
and tension-related extrusion in intestinal villi \cite{Krueger_Clevers_Sciene_2025_Extrusion}.
Line tensions, 
as they could arise from supracellular actin structures, 
facilitate extrusion, 
but do not allow for the extrusion of partially extruded or fully bulged cells as they are observed in some experimental cases.

Our model makes experimental predictions concerning a facilitated extrusion at pentagonal cells and a curvature dependence of the necessary tension decrease for extrusion.
These predictions could be tested in the future in systems such as intestinal organoids in scaffolds \cite{Gjorevski_Science22_tissue_geometry_drives_organoid_patterning},
where the intestinal epithelium is reconstituted by considering an intestinal monolayer on a 3D substrate with similar shape as found in the intestine.
In particular, in this approach cell extrusion has indeed been found to occur at the villus tips \cite{Gjorevski_Science22_tissue_geometry_drives_organoid_patterning}.
It has recently been used to identify tissue tension and cytoskeletal stability to play a role in extrusion \cite{Krueger_Clevers_Sciene_2025_Extrusion} 
and could in the future be used to investigate our proposed dependence on the radius of background curvature.
Recently, also two-dimensional systems to study extrusion have become available \cite{Matjcic_Trepat_2025_Extrusion_2d_organoids} and instabilities could be investigated in such a setting as well.
Intestinal organoids usually display an inverted sign of curvature compared to the villus domain in the intestine with the apical side pointing toward the lumen.
Polarity reversed organoids, however, have the opposite apico-basal orientation and thus the correct curvature. They thus constitute another possible test system to study connections of single-cell shape and tissue curvature \cite{Co_CellRep19_Polarity_reversed_enetroids, Co_NatProtocol21_Polarity_reversal_organoids}.
Although hard to implement in practice, live cell microscopy might be
used to capture extrusion events in more detail and to correlate them with
the topological environment. 
Then it would be interesting to also extend our
theoretical framework to include more dynamic processes, such as cellular rearrangements and tissue fluidity, to achieve a more realistic representation of morphology in dynamic tissues.
Further, more systematic investigations of different extrusion mechanisms, such as basal protrusions (lamellipodia) \cite{Matjcic_Trepat_2025_Extrusion_2d_organoids}, present another interesting avenue for future work.

Our analytical model does not fully describe the morphology of partially extruded cells, because we neglected the relaxation of tilt in the neighboring cells, 
i.e., a non-vanishing angle between the tissue-mid-plane normal and lateral interfaces.
To better capture such intermediate morphologies of cell extrusion, as observed numerically in the BVM, and thus also obtain better predictions of stability, it is necessary to find a continuum description for these states.
One possibility would be to introduce a continuous tilt field, to allow for deviations in the lateral configurations. 
It has been introduced before in the context of lipid membranes \cite{HammKozlov_EPJE2000_Elastic_energy_tilt_bending_membranes, Terzi_JChemPhys17_Tilt_curvature_coupling_lipid_membranes}.
Tilt has been linked to scutoid cell shapes, which should be investigated in this context as well \cite{Lou_PRL23_Scutoids_tilt_curvature_tissues, GomezGalvec_NatComm18_Scutoid_3d_packing_epithelia}.
Together with the framework of a discrete cell in a mean-field tissue, which we presented here, it can serve as a starting point for further investigations of the coupling of single-cell and tissue-scale mechanics.

\begin{acknowledgments}
OMD, MB and USS acknowledge support by the Max Planck School Matter to Life,
with funding by the German Federal Ministry of Education and Research (BMBF),
the Dieter Schwarz Foundation and the Max Planck Society.
MB and USS acknowledge support by the cluster of excellence 3DMM2O (EXC 2082/1-390761711 and EXC 2082/2-390761711)
funded by the Deutsche Forschungsgemeinschaft (DFG, German Research Foundation).
The authors acknowledge the data storage service SDS@hd supported by the Ministry of Science, 
Research and the Arts Baden-W\"urttemberg (MWK) and the DFG through grant INST 35/1503-1 FUGG.
OMD thanks Edouard Hannezo for valuable discussions.
USS is member of the Interdisciplinary Center for Scientific Computing (IWR) at Heidelberg.
\end{acknowledgments}

%
%appendices
%
%
\appendix
\section{Description of \textit{OrganoidChaste}}\label{app:organoidchaste}
The vertex model is implemented in \textit{OrganoidChaste} \cite{OrganoidChaste}, which is a package for Chaste \cite{Chaste_Software_2020}.
Cells are polyhedra with apical, basal and lateral faces, which are described by nodes in 3D, cf.~Fig.~\ref{fig:fig_shells_pentagonal_collapse}(a). 
As nodes of faces are not necessarily coplanar, we interpolate the interfaces by considering an average node for triangulation \cite{Krajnc_PRE18_VM_fluidization, Drozdowski_PRR24_Topological_defects_VM_shells}.
To minimize the energy, Eq.~(\ref{eq:continuum_bubbly_vertex_energy}), the force on node $i$ is calculated based on the energy gradient with respect to the positions of node $\mathbf{r}_i$,
\begin{equation}\label{eq:overdamped_eom_nodes}
    \mathbf{F}_i = - \nabla_{\mathbf{r}_i} E.
\end{equation}
Overdamped dynamics is assumed in the equation of motion of the nodes,
\begin{equation}
    \frac{\mathrm{d}}{\mathrm{d}t} \mathbf{r}_i = \mathbf{F}_i,
\end{equation}
where we assumed a mobility of one.
This equation of motion is solved using Euler stepping.
Volume conservation is achieved by projecting the forces on the space of volume conserving forces via a Lagrange multiplier, corresponding to pressures, with an additional correction step.
 This implementation is identical in our VM code and in the subsequent BVM minimization in Surface Evolver \cite{Brakke_92_Surface_Evolver} described below.

Cells can perform neighbor exchanges, so-called T1 transitions, 
where we first form a protorosette (4-way one-dimensional junction)
and then in a next step create the new interface, following the approach suggested in Ref.~\cite{Krajnc_PRE18_VM_fluidization}.
We perform these randomly with a probability given by a corresponding transition rate on the entire epithelium. 
This has been found to fluidize the 3D VM, allowing for a viscous relaxation of stresses \cite{Krajnc_PRE18_VM_fluidization}. 
In the minimization of the spherical epithelia a decreasing transition rate is taken, similarly to simulated annealing.
The resulting organoid is taken as starting point for a comparative minimization in the VM and the BVM with the same underlying topology.

In order to solve the bubbly vertex model the configuration which is obtained in \textit{OrganoidChaste} is exported to Surface Evolver \cite{Brakke_92_Surface_Evolver}, where the same energy, Eq.~(\ref{eq:continuum_bubbly_vertex_energy}), is considered. 
However, the interpolation points of the triangulation are not longer averaged, allowing for interfacial curvature.
Despite this, the model and the minimization through overdamped relaxation are identical to the VM if no additional effects are added; in particular the energy is unchanged, Eq.~(\ref{eq:continuum_bubbly_vertex_energy}).
If an interfacial bending rigidity is considered, then the mean curvature in Surface Evolver is calculated at triangulation points of the interfaces, which do not belong to the edges of three intersecting interfaces.
This leads to small deviations at the corner triangles due to the neglection of a bending energy across the corresponding interface internal edge (cf.~Fig.~\ref{fig:modulation_pressure_rigidity}(c)), which decreases as the triangulation resolution is increased.

Simulations of the VM were run on the timescale of minutes on a regular computer. 
Minimizations in Surface Evolver for large systems 
were run on a high performance computing cluster on AMD Epyc 7H12 processors, 
with a minimization running on one $2.6\,\mathrm{GHz}$ core for a time from a few hours to up to three days for very large systems, cf.~Fig.~\ref{fig:fig_ico_shells_bubbly}(c).

\section{Experimental methods}
\label{app:organoid_exp_analysis}

Mouse colon (mCol) organoids were generated from a male C57BL/6J mouse as described in Ref.~\cite{Sato2011_Gastroenterology_mcol_organoids}. 
To maintain organoid cultures, mCol organoids were enzymatically and mechanically dissociated into single cells using TrypLE (Gibco). 
Cells were resuspended in Basement Membrane Extract Type R1 (BME-R1) (R\&D Systems), 
embedded in $10\,\mu\text{L}$ domes in wells of 12-well plates (Corning), resulting in 5 domes per well. 
Organoids were cultivated in WENR medium prepared by the addition of $1\times \mathrm{B-27}$ supplement (Gibco), $1.25\,\text{mM}$ n-Acetyl Cysteine (Sigma-Aldrich), $100\,\mathrm{mg}\,\mathrm{mL}^{-1}$ Primocin (InvivoGen), $50\,\mathrm{ng}\,\mathrm{mL}^{-1}$ EGF (Peprotech), $10\%$ (v/v) Noggin Conditioned Medium, $20\%$ (v/v) R-spondin Conditioned Medium, 
$10\,\mu\mathrm{M}$ Y-27632 (TargetMol) and $0.5\,\mathrm{nM}$ Wnt Surrogate (Gibco) into AD3+ basal medium, which consists of Advanced DMEM/F-12 (Gibco) supplemented with $10\,\mathrm{mM}$ HEPES (Sigma‑Aldrich), $2\,\mathrm{mM}$ GlutaMAX (Gibco), and $10\,\mathrm{mM}$ Penicillin/Streptomycin (Sigma‑Aldrich). 
Organoids were cultured at $37^\circ\mathrm{C}$ in a humidified atmosphere with $5\%$ $\mathrm{CO}_2$ (Binder). 
The culture medium was regularly changed every two to three days, and passaging was performed at a 1:6 ratio approximately every five to seven days.

mCol organoids were dissociated into single cells as described above, and seeded at a density of $1000\,\mathrm{cells}\,\mu\mathrm{L}^{-1}$ BME-R1. 
The organoids were embedded in 10 $\mu\mathrm{L}$ BME-1 domes in 1 well of 8-well $\mu$-slide (ibidi), and cultured at $37^\circ \mathrm{C}$ in a humidified atmosphere with $5\%$ $\mathrm{CO}_2$. 
The culture medium was changed on the second day of culture. 
For Latrunculin-A perturbation, organoids were treated for 5 hours with $0.5\,\mu\mathrm{M}$ Latrunculin-A (Cayman Chemical).
In the corresponding paired experiment the wildtype organoids were left in WENR medium.
Following treatment, organoids were fixed ($4\%$ formaldehyde) for 30 minutes at room temperature. $0.5\mu\mathrm{M}$ Latrunculin-A was added to the fixation solution for organoids treated with Latrunculin-A. 
For immunofluorescence staining, organoids were permeabilized in $0.5\%$ Triton-X100 in PBS (Sigma-Aldrich) for 10 minutes at room temperature. 
Organoids were subsequently incubated in blocking buffer ($4\%$ BSA, $0.1\%$ Triton X-100, and $0.1\%$ Tween-20 in PBS) for 1 hour at room temperature. The organoids were then incubated overnight at $4^\circ \mathrm{C}$ with primary antibody (Mouse anti-Beta-Catenin, BD Biosciences, 1:500). 
Organoids were washed with $0.1\%$ Tween-20 in PBS for 3 times before secondary antibody incubation for 1 hour at room temperature (Goat-anti-Mouse-AF488, Life Technologies, 1:500). 
Co-staining for F-actin (Phalloidin-TRITC, Sigma, $50\,\mathrm{ng}\,\mathrm{mL}^{-1}$) and DNA (Hoechst, Life Technologies, $1\,\mu\mathrm{g}\,\mathrm{mL}^{-1}$) was performed simultaneously. 
After three additional washes with $0.1\%$ Tween-20 in PBS, the organoids were immersed in clearing solution ($\mathrm{2.5}\,\mathrm{M}$ fructose in $60\%$ glycerol). 

Images were acquired on a Nikon AX laser scanning confocal microscope using a $25\times$ silicone immersion objective, with an XY pixel size of $0.25\,\mu\mathrm{m}$ and a z-step size of $0.5\,\mu\mathrm{m}$. 
Images were deconvolved using the automatic 3D deconvolution feature in NIS Elements software Version 5.42.03.

\section{Image segmentation}
\label{app:segmentaion_force_inference}

For image segmentation the images were preprocessed by interleaving the 
image stack with averaged images to obtain cubic voxels. 
Further, the pixel intensity was linearly corrected as a function of imaging depth for some organoids, to correct for intensity loss.
Finally, in some organoids gamma correction was performed 
to increase the image contrast.
Using Cellpose 2.0 \cite{Stringer_NatMeth2021_Cellpose1, Pachitariu_NatMeth22_Cellpose2} and ilastik \cite{Berg_NatMeth2019_Ilastik}, 
using the models trained specifically for organoid imaging data with mouse small intestine organoid data in 
Ref.~\cite{Drozdowski_Boonekamp_2025_Force_inference}, 
we segmented the spherical mouse colon organoids.

Using the force inference pipeline we have developed \cite{Drozdowski_Boonekamp_2025_Force_inference}, we then performed three-dimensional reconstruction of the interfaces to obtain normal vectors at the tri-cellular junctions and subsequent force inference.
For this we assume constant surface tensions across cell-cell, cell-lumen or cell-outside interfaces.
This implies the force balance equation at a junction (cf.~Fig.~\ref{fig:exp_mcol}(c))
\begin{equation}
\begin{aligned}
    \Gamma_1 + \Gamma_2 \cos(\theta_{12}) + \Gamma_3 \cos(\theta_{13}) &= 0\\
    \Gamma_2 \sin(\theta_{12}) - \Gamma_3 \sin(\theta_{13}) &=0,
\end{aligned}
\end{equation}
where the three interfaces haves tensions $\Gamma_{1,2,3}$, respectively, and the angle between interface $i$ and $j$ is denoted with $\theta_{ij}$.
Including all interfaces and all junctions yields a large overdetermined system of equations, 
where the angles can be used to infer the underlying (relative) surface tensions $\Gamma$ \cite{Drozdowski_Boonekamp_2025_Force_inference}, with the average tension over all interfaces set to one.
As in Ref.~\cite{Drozdowski_Boonekamp_2025_Force_inference} interfaces which could be associated with segmentation errors were excluded from the analysis.

Both in the BVM and in the experimental data we determine the opening angle of a lateral interface 
by calculating the angle between the apico-basal axis (the vector pointing from the apical (luminal) interface center of mass to the basal (outer) one) 
and the inward pointing lateral interface normal $\varphi$ via $\delta/2 = \pi/2 - \varphi$.
For each cell these lateral opening angles are averaged.
For this calculation interfaces are represented as point clouds.
In the BVM we consider the triangulation points.
For the experimental data we obtain the points by considering voxels that have two cell indices in their 19-point stencil 
(next-to-nearest neighbor voxels) and taking their positions in voxel space.
We determine the lateral normals as the direction with smallest variance via a principal component analysis \cite{Drozdowski_Boonekamp_2025_Force_inference}.

To determine the number of neighbors for each cell, we consider the basal (outer) side, 
as we found segmentation errors to usually occur on the inner apical side.
In the segmentation we sometimes found inclusions of small cells due to segmentation issues \cite{Drozdowski_Boonekamp_2025_Force_inference}.
We thus neglected cells with a volume below $1000$ voxels ($15.625\,\mu\mathrm{m}^{3}$) and then determined tri-interfacial junctions, again via the 19-point stencil.
For each basal interface the number of neighboring junctions, i.e., polygonal edges, was determined as the number of neighbors.
Cells in which segmentation errors were identified (as in the force inference step) are then neglected in the statistics. 
Their basal sides, however, have contributed to the neighbor determination thus not leading to an error from this correction.

\section{Estimating luminal pressure}\label{app:lumen_pressure}
To estimate luminal pressures we consider the vertex model with flat interfaces and its coarse-grained nonlinear stretching energy density \cite{Drozdowski_PRR24_Topological_defects_VM_shells}
\begin{equation}
    e_\mathrm{stretch} = \frac{\Gamma_\mathrm{a}+\Gamma_\mathrm{b}} {\Gamma_\mathrm{l}}\, \Gamma_0 \left( \lambda^2 + 2\frac{1}{\lambda}-3\right),
\end{equation}
with average tension $\Gamma_0=5\, \mathrm{nN}\,\mu\mathrm{m}^{-1}$ and principal stretches in the spherical shell $\lambda=\lambda_\phi=\lambda_\theta$, which describe an elongation $\ell'=(1+\lambda)\ell$ for an in-plane length $\ell$ which is stretched to $\ell'$.
One can show that bending energy contributions constitute higher order corrections which can be neglected.

To obtain the pressure, we consider 
\begin{equation}
    P\,\mathrm{d}[V(R)] = \mathrm{d}[e_\mathrm{stretch}(R)A_0],
\end{equation}
with pressure P, spherical volume $V$ and undeformed spherical area $A_0$; the latter two both depend on the pressured sphere radius $R$.
For a stretched sphere the principal stretch is $\lambda=R/R_0$ with undeformed sphere radius $R_0$.
We can insert the volume and area of a sphere to obtain
\begin{equation}
    P\, 4\pi R^2\,\mathrm{d}R =  4\pi\, \frac{\Gamma_\mathrm{a}+\Gamma_\mathrm{b}} {\Gamma_\mathrm{l}}\,\Gamma_0 \left( 2R - 2\frac{R_0^3}{R^2} \right) \mathrm{d}R,
\end{equation}
which implies for the pressure
\begin{equation}
    P = \frac{\Gamma_\mathrm{a}+\Gamma_\mathrm{b}} {\Gamma_\mathrm{l}}\,\Gamma_0 \left( \frac{2}{R} - 2 \frac{R_0^3}{R^4} \right).
\end{equation}
Note that the first term is the Young-Laplace law which is modified by the second term stemming from the initial shell radius.
For the shell in Fig.~\ref{fig:exp_mcol}(a) with $\Gamma_\mathrm{a}\approx1.02$, $\Gamma_\mathrm{b}\approx1.06$, $\Gamma_\mathrm{l}\approx 0.97$, and $R\approx22\,\mu\mathrm{m}$ we have $R_0= \left(\psi_\mathrm{exp}/\psi_\mathrm{theo}\right)^{1/3}R\approx 18\,\mu\mathrm{m}$, which implies $P\approx440\,\mathrm{Pa}$.

\section{Effect of line tension}\label{app:line_tensions}

\begin{figure}[!b]
	\centering
	\includegraphics[width=\columnwidth]{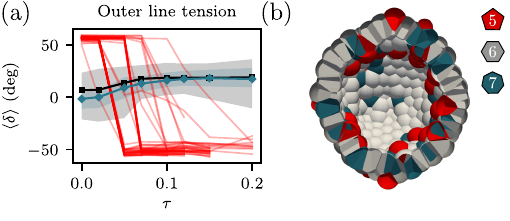}
	\caption{Effect of basal (outer) line tensions in all cells on the opening angles of defect cells. 
    (a) Opening angles as function of the dimensionless line tension $\tau$. 
    Individual red lines mark single pentagonal cells to show single-cell dependency of inward bulging. Symbols show means with shaded regions indicating $5\%$ and $95\%$ quantiles for hexagons in black and heptagons in blue.
    (b) Spherical shell 
    as in Fig.~\ref{fig:fig_shells_pentagonal_collapse} for $\tau=0.05$.}
\label{fig:si_stabilization_linetension}
\end{figure}
\begin{figure}[!t]
	\centering
	\includegraphics[width=\columnwidth]{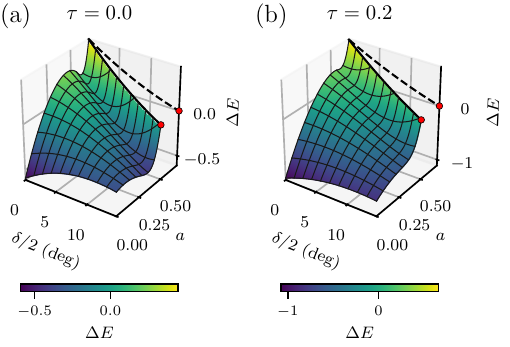}
	\caption{Energy landscapes in configuration space for a pentagonal defect in a tissue with curvature screening and basal line tension $\tau$. The cellular opening angle $\delta/2$ is bounded from above because of the maximum opening angle of the cap, see~Fig.~\ref{fig:critical_tension_defect}(a). Color-coding and lines as in Fig.~\ref{fig:energy_landscapes_extrusion}(a-c). Parameters $\Gamma=0.6$, $R_\mathrm{bg}/\tilde{h}=5$.}
\label{fig:energy_landscapes_linetension}
\end{figure}

The existence of line tensions has been demonstrated in small intestine organoids \cite{Drozdowski_Boonekamp_2025_Force_inference} 
and could in principle stabilize cell shapes basally.
To consider such a process in our model, we add a line tension at all outer basal interface boundaries, which effectively constitutes a contractile contribution along the basal interface edges. 
Fig.~\ref{fig:si_stabilization_linetension} displays the opening angles of BVM simulations with basal (outer) line tensions for all cells, where we see cell-specific variability in switching opening angles through an unbulged intermediate step.
The cell-specific switching again suggests that cell-to-cell variability in tensions is necessary.

Besides a reduction of apico-basal interfacial tension, 
another possible mechanism for extrusion are
contractile actin structures pushing cells outward
\cite{GudipatyRosenblatt_SemCellDevB17_Cell_extrusion_pathways}.
To introduce this, we consider a line tension on the (luminal) apical side.
For a tension $\tau$ we thus have an energy at the defect
\begin{equation}\label{eq:line_tension_energy}
    E_\mathrm{line} = n\, \tau\, a,
\end{equation}
which is an additional $a$-dependent term contributing to the total energy.
It leads to a tilting of the energy landscape in configuration space, 
because in the case of curvature screening no $R_0$ minimization is performed.
Fig.~\ref{fig:energy_landscapes_linetension} displays such energy landscapes for pentagons ($n=5$) for different line tensions $\tau$.
We do see that the tilting can destabilize bulged states.
However, since at $a=0$ the line tension does not contribute anymore, 
an energy barrier in the case of a transitioning from a bulged state with $a=0$ to the extruded case is not changed by the line tension.
In light of partially extruded cells, such as in Fig.~\ref{fig:energy_landscapes_extrusion}(g), 
this implies that an additional tuning of interfacial tensions (either lateral or apico-basal) is necessary to achieve full extrusion.

\section{Details on mean-field tissue parameterization}\label{app:mean_field_energies}

The deformation energy for a tissue parametrized as in Fig.~\ref{fig:extrusion_schematic_defect_cell_in_mean_field_tissue} comprises bending and stretching components, derived in the following. For the cap segment we consider the radial deflection function $f$ with rotational symmetry
\begin{equation}
f(r)=R_0 - \sqrt{R_0^2- (\rho + r)^2},
\end{equation}
where $\rho$ is the radial offset due to the bulged defect cell, cf.~Fig.~\ref{fig:extrusion_schematic_defect_cell_in_mean_field_tissue}.
The bending energy density reads
\begin{equation}
    e_\mathrm{bending} = \frac{\kappa}{2} H^2 + \kappa_\mathrm{G} K,
\end{equation}
with bending rigidity $\kappa$, saddle splay modulus $\kappa_\mathrm{G}$, (total) mean curvature $H=c+c'$, and Gauss curvature $K=cc'$ for principal curvatures $c$, $c'$.
For the deflection function $f(r)$ the curvatures read
\begin{equation}
\begin{aligned}\label{eq:mean_gauss_curvature_cylindrical}
    H(r)&= \frac{1}{r}\partial_r r \left(\frac{\partial_r f}{\sqrt{1+(\partial_r f)^2}}\right)=\frac{2}{R_0} + \frac{\rho}{r R_0},\\
    K(r)&= \frac{\frac{1}{r}(\partial_r f) (\partial_r^2 f)}{\left(1+(\partial_r f)^2\right)^2} =\frac{\rho+r}{r R_0^2}.
\end{aligned}
\end{equation}
With the area element, $\mathrm{d}S = \sqrt{1+{f'}^2}\, r\, \mathrm{d}\phi \mathrm{d}r$, the corresponding bending energy from mean curvature reads

\begin{multline}\label{eq:extrusion_mean_bending_energy_cap}
\frac{E_\mathrm{bending}^\mathrm{mean}}{\kappa \pi} =
- \frac{4}{R_0} \sqrt{R_0^2 - (\rho+r)^2}  - \frac{\rho^2}{\sqrt{R_0^2-\rho^2}} \times \\ \left.  \mathrm{artanh} \left( \frac{R_0^2-\rho\left( \rho + r \right)}{\left(R_0^2- \rho^2\right)^{1/2} \left( R_0^2- \left(\rho +r\right)^2\right)^{1/2}}\right) \right|_{r=r_0}^{r=R_\mathrm{B}}.
\end{multline}
Note that for $\rho=r_0=0$ and small buckling radius $R_\mathrm{B}$, i.e. small $|\nabla f|$ in the outer cone segment, 
the Taylor approximation in $R_\mathrm{B}$ scales quadratically, consistent with the approximation of the total energy with a flat disc as proposed by Seung and Nelson \cite{Seung_88_PRL_Conical_instability_disclination} for the flattened tip for the deformed shape around the defect.
The first term in Eq.~(\ref{eq:extrusion_mean_bending_energy_cap}) arises from the spherical contribution of the mean curvature bending energy, cf. corresponding first term in the mean curvature in Eq.~(\ref{eq:mean_gauss_curvature_cylindrical}). 
The second term, which is proportional to the squared radial offset $\rho^2$, serves as a correction for reducing the radius of the circular segment in cylindrical coordinates, as we decrease the radius of curvature in the azimuthal direction via this offset.
This term stems from the quadratic contribution of $\rho^2$ from the squared mean curvature, Eq.~(\ref{eq:mean_gauss_curvature_cylindrical}), in the bending energy.

The bending energy from Gaussian curvature only depends on the opening angle of the defect cell $\delta$, as this modifies the solid angle covered by the cap segment, consistent with the Gauss-Bonnet theorem,
\begin{equation}\label{eq:extrusion_gauss_bending_energy_cap}
    E_\mathrm{bending}^\mathrm{Gauss} = 2\pi \kappa_\mathrm{G}\left[
    \cos\left(\delta/2\right)-\sin\left(\gamma/2\right) \right],
\end{equation}
with tissue opening angle $\gamma/2$, cf.~Fig.~\ref{fig:extrusion_schematic_defect_cell_in_mean_field_tissue}(a).

The conical part of the bending energy is unchanged to the respective result for large $|\nabla f|$ derived in Ref.~\cite{Seung_88_PRL_Conical_instability_disclination}
\begin{equation}\label{eq:extrusion_bending_energy_cone}
    E_\mathrm{cone} = \frac{f_1}{(1+f_1^2)^{1/2}} \pi \kappa \log(R/R_\mathrm{B}),
\end{equation}
with the total radius $R$, the buckling radius $R_\mathrm{B}$, and the cone slope
\begin{equation}
    f_1^\mathrm{cone} = \left[ \frac{1}{(1-s/2\pi)^2} -1 \right]^{1/2}.
\end{equation}
This slope can be obtained by considering a cone 
where the base's circumference corresponds to the circumference of the corresponding flat disc segment with a wedge of angle $s$ cut-out. 
In the case of a pentagonal defect this implies $s=2\pi/6$.

The stretching energy density reads
\begin{equation}
    e_\mathrm{stretch}= \frac{1}{2} \left( 2\mu \varepsilon_{ij}^2 + \lambda \varepsilon_{kk}^2\right),
\end{equation}
with two-dimensional Lam\'e coefficients $\mu$ and $\lambda$.
To compute the stretching energy of the tissue we consider the in-plane deformation of a conically bent elastic sheet. 
Using polar coordinates $(r,\theta)$, it is clear from symmetry that the azimuthal deformation function must be $u_\theta=q_s\theta r$, because we have a stretch of $q_s r$ 
if we denote with $q_s$ the topological charge (i.e. the relative material fraction of necessary azimuthal stretching). 
For a pentagonal defect this means $q_s=s/2\pi=1/6$, because $1/6$ of the azimuthal circumference is removed. 

Introducing the radial position function $w(r)$, the radial deformation is $u_r(r) = w(r)-r$ and 
the strains for a (moderately) bent plane \cite{Landau_Lifschitz_1970_Book_Theory_Elasticity} in polar coordinates are described by 
\begin{equation}\label{eq:tissue_stretching_von_karman_strains}
\begin{aligned}
    \varepsilon_{rr} &= w'(r)-1+ \frac{(f'(r))^2}{2}, \\
    \varepsilon_{\theta\theta}(r) &= \frac{w(r)-r}{r} + q_s,\\
    \varepsilon_{r\theta}&=0,
\end{aligned}
\end{equation}
where the assumption of moderate bending yields a nonlinear coupling term of the deflection function in the strain.

The energy containing both stretching and bending contributions, c.f.~Eq.~(\ref{eq:total_energy_bending_stretching}), hence constitutes a functional for $w$. The corresponding Euler-Lagrange equation for $w$ is
\begin{equation}\label{eq:extrusion_euler_lagrange_radial_position_w}
    \frac{w}{r} - w' - rw'' + \frac{2\mu}{2\mu+\lambda} \left( q_s-\frac{(f')^2}{2}\right) - rf'f'' = 0.
\end{equation}
Determining the Euler-Lagrange equation for $f$ from the sum of 
bending and stretching energy and considering a cone 
$f'=\mathrm{const.}$ yields the identical result as obtained in Ref.~\cite{Seung_88_PRL_Conical_instability_disclination}, 
implying the bending energy $E=s\kappa \log(R/r_0)$ for some cut-off $r_0$. 
For this result from Ref.~\cite{Seung_88_PRL_Conical_instability_disclination} one uses $w(0)=0$ and $\sigma_{rr}(R)=0$ as boundary conditions.

We now want to take a different approach, however, and consider a ring of material with inner and outer radii $r_0$ and $R_\mathrm{B}$, deflected with a known fixed slope $f'=f_1$. 
Introducing $\Delta s = q_s-{f_1}^2/2$, Eq.~(\ref{eq:extrusion_euler_lagrange_radial_position_w}) is solved by
\begin{equation}\label{eq:extrusion_solution_w_cone_with_slope}
    w(r) = \frac{\mu}{2\mu+\lambda} \Delta s r \log(r) + \frac{k_1}{r} + k_2 r,
\end{equation}
with coefficients $k_1$ and $k_2$, which are determined from the boundary conditions.

Note that the strains in Eq.~(\ref{eq:tissue_stretching_von_karman_strains}) only consider the leading order terms in $f'$.
To also allow for large $f'$, we observe that the azimuthal stretch in a cone is reduced such that for a circle at radius $r$ the deformed radius becomes $r/\sqrt{1+{f_1}^2}$. This results in the corrected radial strain $\varepsilon_{rr}=w'(r)-(1+{f_1}^2)^{-1/2}$.
Since we only consider a linear $f$, i.e. we neglect $f''$, only the term corresponding to $\Delta s$ changes in Eq.~(\ref{eq:extrusion_euler_lagrange_radial_position_w}), implying that the solution is again Eq.~(\ref{eq:extrusion_solution_w_cone_with_slope}), now with
\begin{equation}\label{eq:extrusion_delta_s_with_pentagon}
    \Delta s = q_s - \left( 1- \frac{1}{\sqrt{1+{f_1}^2}}\right).
\end{equation}

\begin{figure}[!t]
	\centering
	\includegraphics[width=\columnwidth]{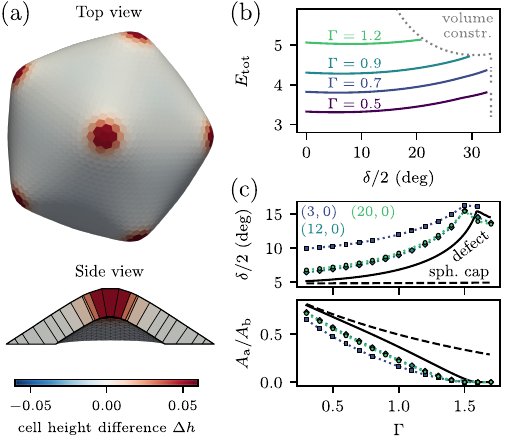}
	\caption{Comparison of continuum model with vertex model simulations of icosahedral shells. 
 (a) VM shell with Caspar-Klug size $(20,0)$ and $\Gamma=1.1$ with top view and side view cross-section. Color-coded is cell height difference from the mean-field VM height (cf.~Eq.~(\ref{eq:equilibrium_cell_height_hex})).
 (b) Total energy of the proposed full defect model as function of defect opening angle $\delta/2$ for different apical/basal surface tensions $\Gamma$. The dotted grey line indicates the angle boundary from the volume constraint. 
 (c) Opening angles $\delta/2$ (top) and apical to basal face area ratios (bottom) for the different models.
 The defect model (solid) considers a defect in a mean-field tissue with the angle $\delta/2$ corresponding to the minimal energy configuration, cf.~(a). The spherical cap model (dashed) corresponds to this energy under the assumption that $\delta/2$ matches the background curvature.
 The symbols indicate vertex model simulation results for different Caspar-Klug indices $(T_1,T_2)$.}
\label{fig:extrusion_comparison_mean_field_VM}
\end{figure}

As boundary conditions we assume vanishing radial stresses at the inner and outer radii, i.e. at $r=r_0$ and $r=R_\mathrm{B}$
\begin{equation}
\sigma_{rr}(r)=2\mu\varepsilon_{rr}(r) + \lambda (\varepsilon_{rr}(r) + \varepsilon_{\theta\theta}(r))=0.
\end{equation}
This boundary condition assumes that no radial (and thus total compressive) stresses act on the boundary of the cap segment and the conical part of the surrounding tissue.
Comparing the cell height in VM simulations to the theoretical mean-field tissue height serves as an estimate for the compressive in-plane stress.
The height difference from the mean-field height (cf.~Appendix~\ref{app:continuum_params}) is shown in Fig.~\ref{fig:extrusion_comparison_mean_field_VM}(a). 

We see that our parameterization for the tissue shape seems reasonable in comparison to the VM shape: around the defect cell we have a ring of tissue which interpolates the opening angle of the cell and the tissue (corresponding to the light grey tissue in Fig.~\ref{fig:extrusion_schematic_defect_cell_in_mean_field_tissue}(a)), surrounded by a conically deformed background tissue (dark grey tissue in Fig.~\ref{fig:extrusion_schematic_defect_cell_in_mean_field_tissue}(a)).
The height difference decays toward the boundary between the cap and the cone segments with a vanishing height difference (and thus stress) in the outer cone segment.
Note, however, that our assumption on the defect cell height does not perfectly match the simulations.
An additional coupling of tissue stress at the inner ring of the surrounding tissue and the defect cell's slant height could be introduced. 
We neglect this additional degree of freedom here in order to obtain a tractable model and as we obtain good agreement to simulations nonetheless.

After determining the coefficients $k_{1,2}$, we integrate the stretching energy to find
\begin{equation}\label{eq:extrusion_stretching_energy_cap}
\begin{aligned}
    \frac{E_\mathrm{stretch}}{\pi Y \left(1+f_1^2\right)^{1/2}} &= \frac{\Delta s^2}{8} \left(R_\mathrm{B}^2 - r_0^2\right) \\
    &\ - \frac{\Delta s^2}{2}\ \frac{R_\mathrm{B}^2 r_0^2}{R_\mathrm{B}^2-r_0^2} \log^2\left(\frac{R_\mathrm{B}}{r_0}\right).
    \end{aligned}
\end{equation}

Coming back to the mean-field tissue, we now approximate the stretching energy in the cap segment (light grey in Fig.~\ref{fig:extrusion_schematic_defect_cell_in_mean_field_tissue}(a)) by replacing it with a cone segment such that the slope matches the opening angle of the defect cell, $f_1=\tan(\delta/2)$.
If we consider the fully extruded case with $r_0=0$ and $\delta=0$, we have the well-known analogous version of a flat disclination for large $|\nabla f|$ \cite{Seung_88_PRL_Conical_instability_disclination}. 
The corresponding approximation for small slopes has been used successfully in the description of viral capsids \cite{Lidmar_PRE_2003_icosahedral_instability_shells}. 
For the conical case, where the full Gaussian curvature is relaxed in the defect cell and $f_1=f_1^\mathrm{cone}$, 
we have $\delta/2=\pi/6$ and $\Delta s=0$, as it should. 
Our approximation thus serves as an interpolation between the two extreme limits of the stretching energy of a flat disclination (as used for viral capsids) and the solution for a cone, depending on the opening angle of the defect cell.

For the outer conical segment of the mean-field tissue (dark grey in Fig.~\ref{fig:extrusion_schematic_defect_cell_in_mean_field_tissue}(a)) we assume that it attains the conical shape and thus the stretching energy vanishes.

Finally, we can write down the full energy of the surrounding tissue as a sum of the contributions in Eqs.~(\ref{eq:extrusion_mean_bending_energy_cap}, \ref{eq:extrusion_gauss_bending_energy_cap}, \ref{eq:extrusion_bending_energy_cone}, \ref{eq:extrusion_stretching_energy_cap}), i.e.
\begin{equation}
    E_\mathrm{tissue} = E_\mathrm{bending}^\mathrm{mean} + E_\mathrm{bending}^\mathrm{Gauss} + E_\mathrm{cone} + E_\mathrm{stretch}.
\end{equation}

\section{Verification of defect cell treatment in icosahedral shells}\label{app:verification_mean_field_ico_shells}

Due to the analytical tractability, icosahedral shells constitute a good test case.
In Ref.~\cite{Drozdowski_PRR24_Topological_defects_VM_shells} we have shown that their shape follows the described theory of moderately bent plates.
Comparing it to simulations, we now show that the above parameterization and continuum theory are able to describe the shape of the pentagonal cell and of the surrounding tissue found in the VM.

The corresponding continuum model energy for this vertex model case consists of the core energy for a defect cell without bulging apical/basal membranes, i.e., with $h_S^a=0=h_S^b$, Eq.~(\ref{eq:extrusion_continuum_bubbly_core_energy}); the bending energy of both the cap segment, Eqs.~(\ref{eq:extrusion_mean_bending_energy_cap}, \ref{eq:extrusion_gauss_bending_energy_cap}), 
and the outer cone region, Eq.~(\ref{eq:extrusion_bending_energy_cone}); and the stretching energy in the cap, Eq.~(\ref{eq:extrusion_stretching_energy_cap}). 
For the resulting total energy $E_\mathrm{tot}$ we assume a large outer radius of the total disc $R$,
such that area changes from deformations in the cone are negligible, and thus drop the term proportional to $\log(R)$ in Eq.~(\ref{eq:extrusion_bending_energy_cone}), as we look at energy differences with respect to a reference state.

In the derivation of the mean field tissue (cf.~Appendix~\ref{app:continuum_params}) lateral interface energies in the energy density are included by counting them half for the respective cell area (to avoid double counting).
Here we now introduce a boundary energy term, which subtracts this lateral energy from the mean field tissue and adds half the pentagonal lattice energy of the defect. 
This lateral energy is present in full and should be considered completely, due to the explicit description of the defect cell.
For the lateral contact between the defect and the tissue we thus subtract the lateral energy contribution which is implicitly given in the mean-field tissue boundary (with a circular cut-out) and replace it by the actual lateral energy contribution of the cell (with a pentagonal shape).
This energy reads
\begin{equation}\label{eq:extrusion_boundary_defect_energy}
    E_\mathrm{bndry} = \frac{5\tilde{h}}{4} \left(2a + \frac{\tilde{h}}{\eta_\mathrm{geom}} \sin(\delta/2)\right) - \pi r_0 \tilde{h}
\end{equation}
and it is added to the total energy, i.e.
\begin{multline}\label{eq:extrusion_sum_all_energies_cone}
    E_\mathrm{tot} = E_\mathrm{core} + E^\mathrm{mean}_\mathrm{bending} + E^\mathrm{Gauss}_\mathrm{bending} + E_\mathrm{cone} \\+ E_\mathrm{stretch} + E_\mathrm{bndry}.
\end{multline}

For the elastic parameters and the bending rigidities of the mean-field tissues we use the mean-field values derived for the VM in Ref.~\cite{Drozdowski_PRR24_Topological_defects_VM_shells}, i.e. Eqs.~(\ref{eq:lame_coefficients_mean_field}, \ref{eq:kappa_kappag_mean_field}) in Appendix~\ref{app:continuum_params}.

We found that for every $\delta/2$ and $a$ the energy has a unique minimum as a function of the buckling radius $R_\mathrm{B}$, or, 
equivalently, as a function of the radial cap curvature radius $R_0$. 
We thus minimize the energy with respect to $R_\mathrm{B}$ to obtain the total energy as a function of the opening angle $\delta/2$ in the vertex model.
Fig.~\ref{fig:extrusion_comparison_mean_field_VM}(b) depicts the energies for different surface tensions $\Gamma$, where we observe a clear minimum. 
Note that for large $\Gamma$ the volume constraint yields an upper bound for the possible opening angles.
Our approach, the defect model, allows the defect to have an opening angle independently from the background tissue curvature.
For comparison we also consider the case of the opening angle following the background curvature, i.e., $\delta/2$ is chosen such that it matches $R_0$ with $\rho=0$, which we henceforth denote as the spherical cap model. 
The offset of the energies is a consequence of the neglection of the $\log(R)$-term and results from the different cell number densities for different $\Gamma$.

In Fig.~\ref{fig:extrusion_comparison_mean_field_VM}(c) we show the opening angle and the ratio of basal and apical interfacial areas in the minimal energy configuration. 
We see that the defect model predicts a strong increase of $\delta/2$ 
until $\Gamma \approx 1.4$ when the basal area completely collapses and the angle is determined by the volume constraint.
The ratio of basal and apical areas, also shows this collapse. 
The symbols indicate vertex model simulations of spheres with different Caspar-Klug indices $(T_1,T_2)$.
Our defect model agrees well with the numerical vertex model data for large enough spheres and captures the shape of the defect 
much better then the model without the defect's ability to adopt a different opening radius as the surrounding tissue.
Note that we do not have any fitting parameters, but rather derive the full model from the vertex model with only one free parameter, namely $\Gamma$.
We thus conclude that the approach of considering a single defect cell in a mean field tissue captures the resulting shape better than an approach purely based on continuum theory, as exemplified by the comparison to the VM here.

\section{Continuum parameters for the mean field tissue}\label{app:continuum_params}

For the continuum description of the VM we consider a flat prototypical hexagonal cell without interfacial curvature, described by the VM energy, Eq.~(\ref{eq:continuum_bubbly_vertex_energy}).
Due to volume conservation, $V=1$, we can relate the hexagonal edge length and the height. Minimizing the energy with respect to height yields equilibrium height \cite{Drozdowski_PRR24_Topological_defects_VM_shells}
\begin{equation}\label{eq:equilibrium_cell_height_hex}
h=2^{1/3}\ 3^{-1/6}\ (\Gamma_\mathrm{a}+\Gamma_\mathrm{b})^{2/3}.
\end{equation}
To derive the elastic stretching parameters, in Ref.~\cite{Drozdowski_PRR24_Topological_defects_VM_shells} a single-cell deformation is considered. 
For the same energy and hexagonal cells the Lam\'e coefficients read
\begin{equation}\label{eq:lame_coefficients_mean_field}
    2\mu = \lambda = (\Gamma_\mathrm{a} + \Gamma_\mathrm{b}),
\end{equation}
which implies for the 2D Young's modulus and Poisson ratio
\begin{equation}\label{eq:youngs_poisson_mean_field}
\begin{aligned}
Y &= \frac{4\mu(\mu+\lambda)}{2\mu+\lambda}  = \frac{3}{2} \left( \Gamma_\mathrm{a}+\Gamma_\mathrm{b} \right),\\
\nu &= \frac{\lambda}{2\mu+\lambda} = \frac{1}{2},
\end{aligned}
\end{equation}
respectively.
To derive the bending rigidity and saddle splay modulus, a curved cell is considered in Ref.~\cite{Drozdowski_PRR24_Topological_defects_VM_shells} and the energy change as function of the cellular curvature for small curvatures is determined to find
\begin{equation}\label{eq:kappa_kappag_mean_field}
\begin{aligned}
\kappa &= \frac{9}{8}\, \frac{1.26}{ 2^{1/3}\, 3^{4/3}} (\Gamma_\mathrm{a} + \Gamma_\mathrm{b})^{1/3},\\
\kappa_{\mathrm{G}} &= \left[ \frac{(\Gamma_\mathrm{a}+\Gamma_\mathrm{b})^2}{2} + \frac{3}{2} - \frac{9}{4} 1.26 \right] \frac{(\Gamma_\mathrm{a} + \Gamma_\mathrm{b})^{1/3}}{2^{1/3} 3^{4/3}}.
\end{aligned}
\end{equation}

In Ref.~\cite{Drozdowski_PRR24_Topological_defects_VM_shells} it has been shown that the VM does not perfectly follow linear elasticity and that the stretching moduli are functions of the strains.
Thus a nonlinearity correction for the icosahedral instability is necessary when considering the continuum limit of the VM, i.e. $k_\mathrm{ico}=1/2$.

%\bibliographystyle{apsrev4-2}
%\bibliography{references}{}

%apsrev4-2.bst 2019-01-14 (MD) hand-edited version of apsrev4-1.bst
%Control: key (0)
%Control: author (8) initials jnrlst
%Control: editor formatted (1) identically to author
%Control: production of article title (0) allowed
%Control: page (0) single
%Control: year (1) truncated
%Control: production of eprint (0) enabled
%

\end{document}